\documentclass[final,journal]{IEEEtran}

\usepackage[pdftex]{graphicx}
\DeclareGraphicsExtensions{.eps,.jpg}
\usepackage{amssymb}
\usepackage{comment,cite}
\usepackage{algorithm,algcompatible}
\usepackage{color}
\usepackage{multirow}
\usepackage{wrapfig}
\usepackage{bm,soul}
\usepackage{url}
\usepackage{epstopdf}
\ifCLASSOPTIONcompsoc
\usepackage[caption=false,font=normalsize,labelfont=sf,textfont=sf]{subfig}
\else
\usepackage[caption=false,font=footnotesize]{subfig}
\fi

\usepackage[acronym,shortcuts]{glossaries}

\DeclareMathOperator*{\argmax}{arg\,max}

\newacronym{FA}{FA}{false alarm}
\newacronym{iid}{iid}{independent and identically distributed}
\newacronym{KL}{KL}{Kullback-Leibler}
\newacronym{LLR}{LLR}{log-likelihood ratio}
\newacronym{MD}{MD}{missed detection}
\newacronym{PDF}{PDF}{probability density function}
\newacronym{CDF}{CDF}{cumulative density function}
\newacronym{CCDF}{CCDF}{complementary CDF}
\newacronym{MSE}{MSE}{mean square error}
\newacronym{OFDM}{OFDM}{orthogonal frequency division multiplexing}
\newacronym{SINR}{SINR}{Signal-to-Interference-and-Noise Ratio}

\makeatletter
\def\ve#1{{\mathchoice{\mbox{\boldmath$\displaystyle #1$}}%
              {\mbox{\boldmath$\textstyle #1$}}%
              {\mbox{\boldmath$\scriptstyle #1$}}%
              {\mbox{\boldmath$\scriptscriptstyle #1$}}}}
\makeatother

\makeatletter
\newcommand\remembertext[2]{
  \immediate\write\@auxout{\unexpanded{\global\long\@namedef{mytext@#1}{#2}}}%
  {#2}%
}
\newcommand\recalltext[1]{%
  \ifcsname mytext@#1\endcsname
    \@nameuse{mytext@#1}%
  \else
    ``???''
  \fi
}
\makeatother

\graphicspath{{./figures/}}

\begin{document}

\title{Cooperative Authentication in Underwater\\Acoustic Sensor Networks}
\author{\IEEEauthorblockN{Roee~Diamant,~\IEEEmembership{Senior~Member,~IEEE}, Paolo~Casari,~\IEEEmembership{Senior~Member,~IEEE}, Stefano~Tomasin,~\IEEEmembership{Senior~Member,~IEEE}}%
    \thanks{Manuscript received February 21, 2018; revised August 14, 2018; accepted December 10, 2018. The associate editor coordinating the review of this paper and approving it for publication was S. Yang. (\emph{Corresponding author: Roee Diamant.)}. This research has been sponsored in part by the NATO Science for Peace and Security Programme under grant G5293.}
    \thanks{R. Diamant is with the Department of Marine Technologies, University of Haifa, Haifa 3498838, Israel (e-mail: roeed@univ.haifa.ac.il).}%
    \thanks{P. Casari is with the IMDEA Networks Institute, 28918 Madrid, Spain (e-mail: paolo.casari@imdea.org).}%
    \thanks{S. Tomasin is with the Department of Information Engineering, University of Padova, 35122 Padova, Italy (e-mail: tomasin@dei.unipd.it).}%
    \thanks{Color versions of one or more of the figures in this paper are available online at http://ieeexplore.ieee.org.}%
    \thanks{Digital Object Identifier 10.1109/TWC.2018.2886896}%
}

\maketitle

\begin{abstract}
With the growing use of underwater acoustic communications (UWAC) for both industrial and military operations, there is a need to ensure communication security. A particular challenge is represented by underwater acoustic networks (UWANs), which are often left unattended over long periods of time. Currently, due to physical and performance limitations, UWAC packets rarely include encryption, leaving the UWAN exposed to external attacks faking legitimate messages. In this paper, we propose a new algorithm for message authentication in a UWAN setting. We begin by observing that, due to the strong spatial dependency of the underwater acoustic channel, an attacker can attempt to mimic the channel associated with the legitimate transmitter only for a small set of receivers, typically just for a single one. Taking this into account, our scheme relies on trusted nodes that independently help a sink node in the authentication process. For each incoming packet, the sink fuses beliefs evaluated by the trusted nodes to reach an authentication decision. These beliefs are based on estimated statistical channel parameters, chosen to be the most sensitive to the transmitter-receiver displacement. Our simulation results show accurate identification of an attacker's packet. We also report results from a sea experiment demonstrating the effectiveness of our approach. 
\end{abstract}


\section{Introduction}\label{sec:Intro}

Underwater acoustic networks (UWANs) are increasingly being perceived as a cost-effective means of ocean exploration and monitoring. While carrying out these tasks, however, UWANs are left unattended over long periods of time, and may become vulnerable to external attacks. The recent introduction of standards for underwater acoustic communications (UWAC)~\cite{Janus} makes these attacks more probable. For this reason, the investigation of security mechanisms tailored to the specific UWAC characteristics is currently gaining momentum. In particular, it has been observed that only in very specific cases the same security techniques developed for terrestrial wireless radio networks can be directly applicable in underwater scenarios. For example, this is the case of well-known elliptic-curve cryptography schemes, which have been evaluated for underwater applications in~\cite{souza_e2e_2013}. In most other cases, a systematic re-thinking of security schemes and strategies has to be carried out~\cite{han_jiang_secure_comm_uw_commag_2015}. 

The focus of this paper is the authentication of a packet received by a sink node with the support of trusted nodes. Authentication mechanisms allow a node to prove that it is a legitimate member of a network, so that controller nodes or sinks can trust the data sent by the node. This step is of great importance, especially in underwater monitoring tasks and tactical scenarios. The sink's objective is to determine whether the packet is coming from either the legitimate node or the attacker. Conversely, the attacker's objective is to let its own packet be recognized as authentic by the sink. The authentication process is based on the acoustic communication channel features, rather than on cryptographic techniques. Trusted nodes cooperate to the authentication process without knowing its outcome: each trusted node independently sends authentication data to the sink, which does not broadcast the authenticity decision, in order to increase the system's spectral efficiency and avoid additional security risks. We remark that the trusted nodes do not need to exchange data in order to complete the authentication process.

Our approach is based on channel features that mildly vary over time and space, slow enough such that their distribution can be approximated as constant during the authentication process. We choose these features by examining a data set from more than a hundred sea experiments, and by showing that the distributions of different channel features --such as the number of channel taps, the relative delay spread, and the received power level-- are sufficiently stable over time and sufficiently diverse for different transmitter-receiver pairs. This makes such features amenable for authentication purposes.  
For incoming packets received by different trusted nodes, we leverage this diversity by systematically measuring the distribution of the evaluated channel's characteristics, and by calculating beliefs that are then transmitted to and processed by a sink node. The outcome is a measure to discriminate between packets sent by a legitimate transmitter and packets received from an attacker. We have tested the performance of our authentication method both in simulations based on established acoustic propagation models and in two sea experiments. A comparison with existing approaches and a theoretical bound are also provided. The results show a very good trade-off between the probability of detecting an attack and the probability of a false alarm. 

In the following, we survey the state of the art in UWAC authentication (Section~\ref{sec:survey}), and describe our system model, including the features we use for authentication (Section~\ref{sec:model}). We discuss our authentication method (Section~\ref{sec:methods}) and evaluate it through simulations (Section~\ref{sec:numres}) and two sea experiments (Section~\ref{sec:exp}), before drawing conclusions in Section~\ref{sec:conc}.

\section{Related Work}\label{sec:survey}

Comprehensive reviews of security approaches for underwater networks are provided in~\cite{han_jiang_secure_comm_uw_commag_2015,petroccia_conti_directions_2016}, which discuss approaches for anti-jamming, privacy securing, and covertness. As this paper focuses on authentication, in this section we consider only authentication approaches. In~\cite{souza_e2e_2013}, an end-to-end authentication scheme has been proposed using the elliptic curve digital signature algorithm (ECDSA). Secret keys can also be used for authentication, and in this respect 
a robust secret key generation scheme is proposed in~\cite{liu_robust_key_generation_2008}. 
The survey in~\cite{Dalhatu2018} covers several forms of cooperation and their application to underwater networks. Game theory is advocated as a means to foster cooperation among the nodes. In a security context, this concept is applied to motivate nodes to behave cooperatively, and to improve the effectiveness of end-to-end authentication schemes, which are seen as an important aspect of future underwater network developments~\cite{Sharif-Yazd2017}.

User authentication has been addressed by the so-called physical layer security by exploiting the channel coherence, i.e., messages going from the same source to the same destination are subject to the same channel, whereas a message coming from a fake source located in a different position will be subject to a different channel, which can be estimated at the destination (see~\cite{7270404} for a survey). 
Various channel features have been considered in this context. For example, in wireless radio systems, the power level was considered in~\cite{Faria,demirbas_06,Chen}, the impulse response of a wideband channel in~\cite{xiao_07,xiao_08,Xiao09,Xiao10,He-mil09}, the frequency response of an \ac{OFDM} transmission in~\cite{He-icc09}, the power spectral densities in~\cite{6584938},
and the time series of the received signal strength in a data/acknowledgment packet exchange in~\cite{Qiuhua2017}.

When multiple nodes are available, the authentication process can be further enhanced. 
In~\cite{laurenti_tomasin_anchor_selection_2016}, a distributed authentication in wireless networks was considered where multiple sensors report their correlated measurements to a fusion center, which makes the ultimate authentication decision. In~\cite{8108524}, a similar solution using compressed sensing is studied. All these works typically assume a given channel statistic, which is used to formulate the authentication hypothesis testing problem. In~\cite{8234675}, this assumption is removed by using logistic regression techniques, which is applied to the received signal power.

Different from the literature presented so far, our authentication approach relies on the estimation of the \emph{distribution} of selected underwater channel features instead of their instantaneous realizations, as opposed to, e.g.,~\cite{Faria,demirbas_06,Chen,xiao_07,xiao_08,Xiao09,Xiao10,He-mil09,He-icc09,6584938,Qiuhua2017}. Unlike~\cite{liu_robust_key_generation_2008}, we do not explicitly generate keys starting from channel features; we do not base our scheme on cryptography (unlike~\cite{souza_e2e_2013}); and in order to save energy we do not require the transmission of structured jamming signals, unlike some approaches in~\cite{han_jiang_secure_comm_uw_commag_2015}. Moreover, most of the existing literature on physical layer authentication is based on the comparison of a reference channel with the estimated channel, in a \ac{MSE} sense. This approach is optimal when the estimate is only subject to Gaussian noise but is not adequate in a UWAC context, where instantaneous channel variations and noise requires a more elaborate description of the PDF of the channel estimate (and its variations with respect to the reference channel). Therefore, we derive the distribution of the channel features directly from packets received by trusted nodes, and fuse their beliefs at a sink to make a decision on the authenticity of the message. Our approach proves valid both in simulations, where we realistically model the underwater acoustic channels via ray tracing~\cite{bellhop}, and in a sea experiment, proving that the assumptions behind our scheme are valid in practice.

\section{System Model}\label{sec:model}

We consider an underwater acoustic network with a sink node assisted by $N$ trusted nodes, one legitimate transmitter, and one attacker. The extension to deployments consisting of multiple attackers is straightforward. Transmissions are organized in packets of $T$ symbols. We assume that the first
packet always comes from the legitimate node. Then, the following packets can arrive from the legitimate transmitter or from the attacker. Each packet is labeled with a unique identification (ID) number that prevents replay attacks. Moreover, the trusted nodes are assumed to be roughly time-synchronized 
such that the times at which the trusted nodes receive a packet with a specific ID (according to their own clock) are within a reasonable time span of up to one maximum propagation delay in the network. This prevents the retransmission of the same packet by the attacker with beamforming/channel modification techniques aimed at bypassing the authentication procedure.

\subsection{Assumptions about the channel features} \label{sec:assumptionsfeat}

In the UWAC channel, the transmitted signal is reflected both by the sea boundaries and by volume scatterers, such as plankton and sediments. Moreover, due to the continual motion of the waves, signals are affected by a Doppler shift varying up to tens of Hz~\cite{Doppler2}. The channel is modeled as a tapped delay line, whose delay spread is in the order of hundreds of milliseconds~\cite{stojanovic2009underwater}, with significant variations for different transmitter-receiver locations. The power attenuation of the UWAC channel is governed by the propagation loss and the absorption loss. The propagation loss is a function of the channel structure and entails a highly non-linear process~\cite{Acoustics}.
Moreover, the sound propagation between layers, characterized by different sound speed (mostly due to water pressure or temperature changes), leads to both refraction and additional attenuation. The absorption loss is a function of the carrier frequency, and is mainly affected by water temperature, salinity, and pressure~\cite{Acoustics}. Since propagation and absorption losses cannot be separated at the receiver, we consider the overall attenuation level.

Our proposed authentication method relies on features extracted from the underwater acoustic channel, such as the number of channel taps, the channel's delay spread, and the received power level. These features are typically location-dependent, such that the channel realization observed by a trusted node for packets received from the legitimate node is different than that observed by an attacker located far from the legitimate node. Even if underwater channels are time-varying, we assume that the feature statistics do not change, as long as the locations of the transmitter and receiver remain relatively stable. In Section~\ref{sec:statchanfeatanalysis}, we will validate this assumption based on more than 100 sea experiments conducted at various times and in diverse environments over a four-year period.

Let ${\cal I}$ be the set of channel features used for authentication, and $x_{i,n}(t)$, $i \in {\cal I}$, be the estimated channel feature extracted at time $t \in \mathcal{T}_\phi$ at trusted node $n \in [1, N]$, where $\mathcal{T}_\phi$ is the set of time epochs when a feature $x_{i,n}(t)$ has been estimated for packet $\phi$.
Each random variable $x_{i,n}(t)$ is modeled as a generalized Gaussian (GG), whose statistics is determined by the three parameters
\begin{equation}
\omega_{i,n} = (\mu_{i,n}, \sigma_{i,n}, \beta_{i,n})\;,
\label{e:distribution2}
\end{equation}
where $\mu_{i,n} = \mathbb E[x_{i,n}(t)]$ is the mean, $\mathbb E[x_{i,n}^2(t) - \mu_{i,n}^2] = \frac{\sigma_{i,n}^2 \Gamma(3/\beta_{i,n})}{\Gamma(1/\beta_{i,n})}$ is the variance, and $\beta_{i,n}$ is a shape parameter. The GG distribution is chosen to provide flexibility. In particular, it models the Laplace distribution ($\beta_{i,n}=1$), the Gaussian distribution ($\beta_{i,n}=2$) and the uniform distribution ($\beta_{i,n}\rightarrow\infty$). We will provide more details on how to obtain these parameters for authentication purposes both in Section~\ref{sec:hypo} and in the Appendix. Then the conditional \ac{PDF} of $x_{i,n}(t)$ (for a specific set of parameters $\omega_{i,n}$) is~\cite{Novey:2010}
\begin{equation}
p_{x| \omega}(a|\omega_{i,n}) =\frac{\beta_{i,n}}{2\sigma_{i,n}\Gamma\big(\frac{1}{\beta_{i,n}}\big)}{\rm e}^{-\big(\frac{|a-\mu_{i,n}|}{\sigma_{i,n}}\big)^{\beta_{i,n}}}\;.
\label{e:distribution}
\end{equation}

As anticipated above, we assume that the feature statistics $\omega_{i,n}$ are static during the authentication process, i.e., they do not change over time, or across different packets for fixed nominal locations of the transmitter and receiver. The validity of this assumption is confirmed by more than one hundred sea experiments, as described in Section~\ref{sec:param}. We also carried out two sea experiments (Section~\ref{sec:exp}) where we successfully tested the authentication scheme. This further confirms that the stability assumption is appropriate in practice.

\subsection{Features for physical layer authentication}\label{sec:param}
 
The key features we use to determine the integrity of an underwater acoustic transmission are the number of channel taps, the tap power, the coherence time, the root mean square (RMS) of the relative delay experienced by the channel taps with respect to the first tap, and a smoothed version of the received power. The effectiveness of the authentication features is provided by the fact that they are relatively stable over time, so their distribution can be approximated as static, but varying considerably in space, so that an attacker located in a different position can be distinguished from the authentic transmitter.
In this section, we discuss the rationale behind our choice of channel features.

Consider a measured channel power-delay profile $H_n(t, \tau)$ for time $t$, delay $\tau$, and node $n$. Define the relevant tap set ${\mathcal S}_n(t) = \{\tau: |H_n(t,\tau)|> T_h\}$, where $T_h$ is a threshold chosen to achieve a probability of false alarm of $10^{-4}$ in the identification of relevant channel taps at the output of a normalized matched filter, according to the analysis in~\cite{MF}.
The following channel features are considered for authentication by trusted node $n$:

\noindent {\bf 1---Number of channel taps.}
The estimated number of significant taps is 
\begin{equation}
x_{1,n}(t) = |{\mathcal S}_n(t)|\;.
\label{e:TapNum}
\end{equation}
While the number of channel taps could be limited or location-independent in specific scenarios (e.g., deep water), there exists a broad class of shallow-water scenarios where the number of taps can vary widely, depending not only on the location of the transmitter and receiver, but also on their depth. For this reason, we consider the number of taps a relevant authentication feature.

\noindent {\bf 2---Average tap power.}
From the power-delay profile $H_n(t, \tau)$, we measure this metric as
\begin{equation}
x_{2,n}(t) = \frac{1}{|{\mathcal S}_n(t)|}\sum\limits_{\tau\in {\mathcal S}_n(t)}|H_n(t,\tau)|\;.
\label{e:TapPower}
\end{equation}%

\noindent {\bf 3---Coherence Time.}
This metric conveys the amount of time when the channel can be approximated as stationary. Formally, we define
\begin{equation}
x_{3,n}(t) \!=\! \mathrm{argmax}(\Delta) \ s.t. \ \mathrm{NC}\big(H_n(t,\tau), H_n(t-\Delta,\tau)\big)\!>\!0.9\;,
\label{e:Coherence}
\end{equation}
where $\mathrm{NC}\left(\alpha,\beta\right)$ is the normalized correlation between $\alpha$ and $\beta$.

\noindent{\bf 4---Relative RMS delay spread.}
Call $\tau_0 = \min \{\tau : \tau \in {\cal S}_n(t)\}$ the delay of the first tap. The estimate of the relative RMS delay spread is computed as
\begin{equation}
x_{4,n}(t) = \bigg( \frac{1}{|{\cal S}_n(t)|-1}\sum_{\tau \in \mathcal {\cal S}_n(t), \tau\neq\tau_0} (\tau-\tau_0)^2 \bigg)^{1/2} \;.
\label{e:Delay}
\end{equation}

\noindent {\bf 5---Average path delay.}
The average path arrival delay, relative to the arrival of the first tap, is
\begin{equation}
x_{5,n}(t) = \frac{1}{|{\mathcal S}_n(t)|-1}\sum\limits_{\tau\in {\mathcal S}_n(t), \tau\neq\tau_0}(\tau-\tau_0)\;.
\label{e:PathDelay}
\end{equation}

\noindent {\bf 6---Smoothed received power.}
Since we expect the received power to change, we consider the difference between the current power measurement and a smoothed version of the previous measurements. Formally, let $q_{n,t}$ be the  power of a symbol received at time $t$ by node $n$, call $t'$ the time when the previous feature measurement was performed, and let $0 \leq \alpha \leq 1$ be a user-defined parameter. The smoothed received power is recursively defined as
\begin{equation}
x_{6,n}(t)=\alpha\,q_{n,t} + (1-\alpha)\, x_{6,n}(t')\;.
\label{e:smooth}
\end{equation}

Other channel features can also be used for authentication. For example, the Doppler shift would seem a good candidate, since it inherently changes for different transmitter/receiver locations. However, relying on the Doppler shift for authentication requires the legitimate node to notify others of its mobility plans. In addition, the attacker can easily impress a given Doppler shift on its packets by resampling its transmitted signals. These aspects make Doppler shift a comparatively less robust channel feature, hence in this work we do not consider it for authentication.
Similarly, the time of arrival of a packet transmission would also be a good channel feature for authentication purposes. However, the trusted node would need to cooperate with the transmitter in order to determine the transmission time: this makes the time of arrival vulnerable to attacks, and therefore we avoid using it. 

\begin{figure}[t]
       \centering
\includegraphics[width=8cm]{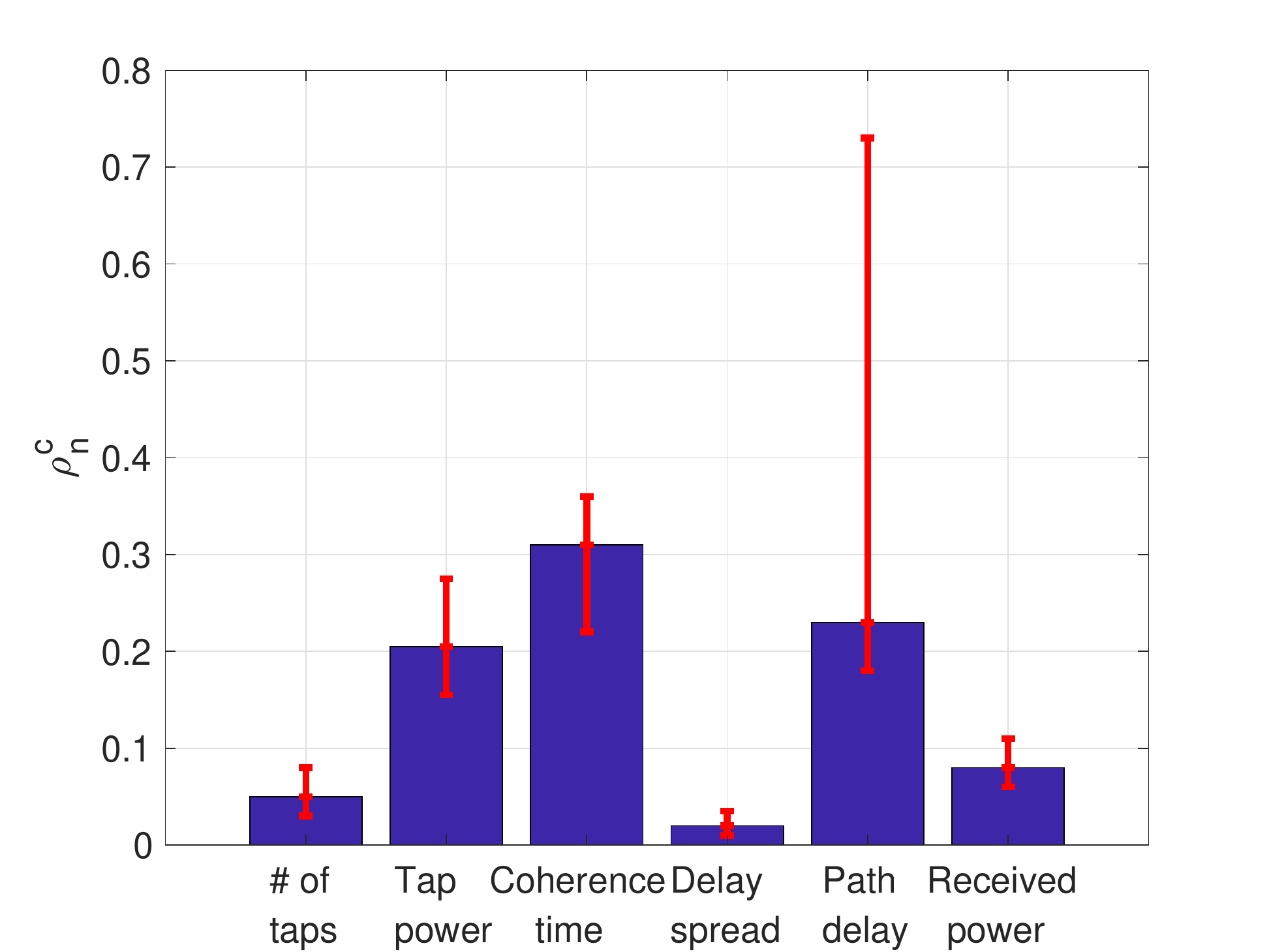}
       \caption{Spatial and time dependency of some acoustic channel features: the red line on each histogram element conveys the 10th and the 90th percentile for each metric.}
       \label{f:expexample1}
\end{figure}

\subsection{Statistical analysis of the channel features} \label{sec:statchanfeatanalysis}

In order to assess the sensitivity of the channel features to the transmitter and receiver locations, we performed a statistical analysis using time-based channel impulse responses recorded during more than 100 sea experiments at different locations, water depths, and times of the year. All experiments were performed along the Israeli coast between 2001 and 2005, and included the use of various acoustic equipment at different carrier frequencies. The channels were recorded for at least~10~s with a resolution of~50~ms. The database format is reported in~\cite{Diamant:2005}. 
 
From the database in~\cite{Diamant:2005}, we selected the experiments that included more than two nodes, so that we could compare the differences between the considered channel features in space at the same time. To capture fluctuations due to environmental changes on a small time scale, this comparison is performed for the time series rather than for average values. As the receivers used similar hydrophones for reception, in this section we drop the node index $n$ from the notation. 

For the same transmitter and a pair of receivers $i$ and $j$ within the set of receiver pairs ${\cal R}$ in the same sea experiment, we evaluate the difference between the time series of the estimated channel features $\ve{x}_{j}$ and $\ve{x}_{i}$ with elements $x_{j}(t)$ and $x_{i}(t)$, respectively, obtained at the same time instances $t=0,\ldots,T_N-1$.  The difference between the two time series provides a measure of the spatial dependency of the estimated feature. Yet, to be used as authentication metrics, the chosen features must also not change significantly during the acquisition time. As a metric for the comparison, we used
\begin{eqnarray}
\rho^c=\frac{1}{|{\cal R}|}\sum_{(i,j) \in{\cal R}\, , i\neq j}
\frac{\sum_{t=0}^{T_N-1} x_j(t)x_i(t)}{\sqrt{\sum_{t=0}^{T_N-1}x_j^2(t)\sum_{t=0}^{T_N-1} x_i^2(t)}} \cdot \nonumber\\
\hspace{2cm}\times \left(1-\rho^{\mathrm{diff}}(\ve{x}_{j})\right)\left(1-\rho^{\mathrm{diff}}(\ve{x}_{i})\right)\;,
\label{e:compare}
\end{eqnarray}%
where $|\cdot|$ is the cardinality of $\cdot$, and $\rho^{\mathrm{diff}}(\ve{x}_{j})$ is Jain's fairness index~\cite{Jain:1984}
\begin{equation}
\rho^{\mathrm{diff}}(\bm{x}_j)= \bigg(\sum_{t=0}^{T_N-1}x_j(t)\bigg)^2 \bigg(T_N \sum_{t=0}^{T_N-1} x_j^2(t)\bigg)^{-1}\;,
    \label{e:fair}
\end{equation}
used to estimate the variation of the considered channel feature over time. 
The value of $\rho^c$ decreases if a given metric varies as a function of a node's location, but remains otherwise stable over time. Hence, for authentication, those features yielding a small $\rho^c$ are preferred. 

\begin{figure}[t]
       \centering
       \subfloat[Receiver 1\label{f:Hist_Numtaps1}]{{\includegraphics[width=.85\columnwidth]{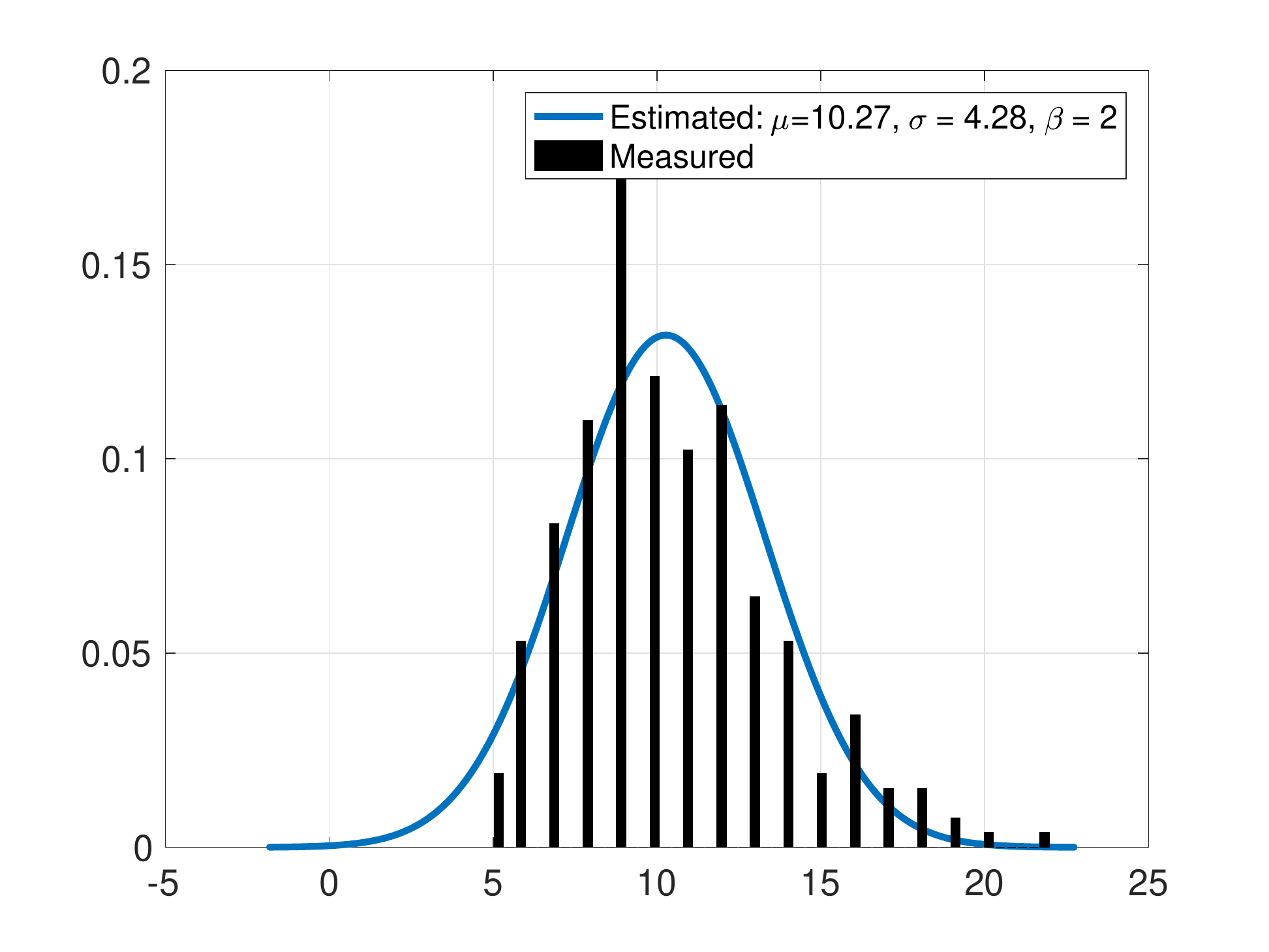}}}\\
       \subfloat[Receiver 2\label{f:Hist_Numtaps2}]{{\includegraphics[width=.85\columnwidth]{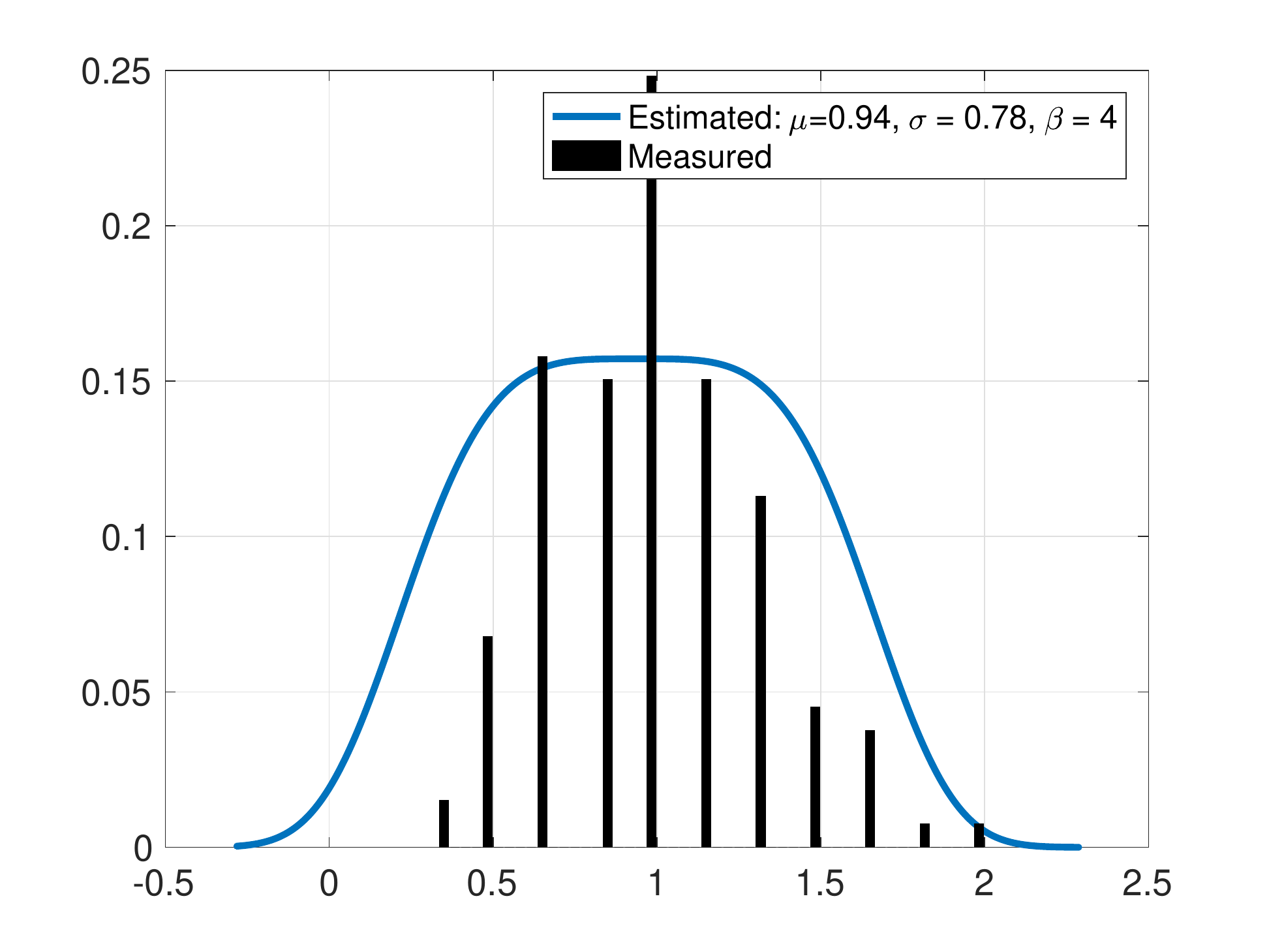}}}
       \caption{Empirical and theoretical pdf of the number of channel taps from one of our sea experiments.}
       \label{f:expexample2}
\end{figure}
Fig.~\ref{f:expexample1} shows the average $\rho^c$ for the different channel features considered in Section~\ref{sec:param}. All features were measured by estimating the channel impulse response over time (see full description in~\cite{Diamant:2005}), and by processing it to extract the feature values. While there is a considerable difference between features estimated at different locations, we observe that the most significant differences are obtained for the number of taps and the relative delay spread, and that such differences hold in a variety of experiments, as can be inferred from their small confidence intervals. 

Finally, the results confirm that the flexibility offered by the generalized Gaussian distribution in~\eqref{e:distribution} is indeed required in practical settings. Figs.~\ref{f:Hist_Numtaps1} and~\ref{f:Hist_Numtaps2} show the empirical and theoretical PDFs of the number of taps, estimated in our sea experiment, as observed by two receivers. We observe that the shape of the distributions is different for each receiver. This demonstrates why obtaining a specific distribution for the channel is challenging. Still, due to its flexibility, our estimated GG distribution~\eqref{e:distribution} is able to capture the empirical PDF quit well.

\subsection{Assumptions on the Attacker Model} \label{sec:attackermodel}

We assume that the attacker is a single malicious node, able to perform any kind of signal processing on its transmitted packet. In particular, it has unlimited transmission power capabilities, it can filter the transmitted signal in order to let the trusted node estimate a different channel, and it can superimpose signals onto the message. 
We also assume that the attacker knows:
\begin{enumerate}
\item the position of both the trusted nodes and the legitimate node;
\item the training signals used for channel estimation by the trusted nodes;
\item the instantaneous acoustic channel realization observed over the link from itself to all of the trusted nodes plus the legitimate node;
\item the statistical correlation of the above channels with the channel realizations observed over the links from the legitimate node to each of the trusted nodes.
\end{enumerate}

\section{Authentication Method}\label{sec:methods}

In this section, we outline the details of our authentication method. Our scheme aims at establishing the authenticity of a received packet. This is a hypothesis-testing problem, where for a packet $\phi$ the two hypotheses are 
\begin{itemize}
\item  $\mathcal H_0$: packet $\phi$ is authentic, and
\item $\mathcal H_1$:  packet $\phi$ is not authentic.
\end{itemize}
By \ac{FA} we mean the case where an authentic packet is classified as not authentic; by \ac{MD} the case where a non-authentic packet is classified as authentic. For a given testing strategy, the \ac{FA} and \ac{MD} probabilities are the probabilities that a \ac{FA} or a \ac{MD} occur.

Our authentication is based on the set of features $\bm{X}(\phi) = \{x_{i,n}(t) , t \in \mathcal{T}_\phi, \forall i,n\}$  associated with packet $\phi$. The decision on $\bm{X}(\phi)$ is made using alternative distributions with parameters estimated from previously received packets. Let $p_{\bm{X}|\mathcal H}(\bm{X}(\phi)| \mathcal H_m)$ be the joint \ac{PDF} of $\bm{X}(\phi)$ conditional upon hypothesis $\mathcal H_m$. Then, for packet~$\phi$, the optimal hypothesis testing strategy is obtained by using the \ac{LLR}~\cite{Mudd_EM} as decision index:
\begin{equation}
\Psi_\phi = \Psi_\phi^0 - \Psi_\phi^1\;, 
\label{e:psiphi}
\end{equation}
where ($m =\{0,1\}$)
\begin{equation}
\Psi_\phi^m = \log p_{\bm{X}|\mathcal H}\big(\bm{X}(\phi)|\mathcal H_m\big)\;.
\label{e:psi}
\end{equation}
The decision procedure for the hypothesis testing problem therefore becomes
\begin{equation}
\label{e:threshold}
\widehat{\mathcal H} = \begin{cases}
\mathcal H_0 & \Psi_{\phi} < \lambda\,, \\
\mathcal H_1 & \Psi_{\phi} \geq \lambda\,, \\
\end{cases}
\end{equation}
where $\lambda$ is a threshold that trades off between the \ac{FA} and \ac{MD} probabilities. 

\subsection{Distributed Authentication}  \label{sec:distribauth}

Our authentication method is distributed, as each trusted node pre-processes each received packet to reduce the amount of data exchanged with the sink node. In turn, the sink will determine the authenticity of each packet.%
\footnote{We remark that this requires both error control (to ensure reliable transmission from the trusted nodes to the sink) and medium access control (to regulate the access of the trusted nodes to the UWAC channel). Since the amount of information exchanged from the trusted nodes to the sink is small and limited to the value $R_{\phi, n}$ in~\eqref{e:psi2b}, the above controls can be implemented using reliable communication schemes~\cite{mfsk_dsss_comparison_2014}. \label{fn:outside}} %
In the following, we derive both the data provided by the trusted nodes and the fusion procedure at the sink.

We assume that a set $\cal I$ of channel features has been preliminarily selected to be used for authentication purposes. In fact, the chosen set of features matches the characteristics of a given environment, and is usually designed to extract the maximum diversity from that environment.
We consider~\eqref{e:distribution} as the \ac{PDF} of the channel features used for authentication; however, we do not know the parameters $\omega_{i,n}^m$ a priori. Instead,
we estimate $\omega_{i,n}^m$ using all received packets up to packet $\phi$, and we indicate this estimate as 
$\hat{\omega}^m_{i,n}(\phi) = \big(\hat{\mu}_{i,n}(\phi), \hat{\sigma}_{i,n}(\phi), \hat{\beta}_{i,n}(\phi)\big)$.

As analyzed in \cite{Porter:1987} and discussed in \cite{UWAC}, we rely on the spatial diversity of the UWAC channel to assume that channel variations from the transmitter to each of the trusted nodes are independent. Under this assumption, we use the estimated parameters to compute the conditional probability ($m=\{0,1\}$)
\begin{equation}
p_{\bm{X}|\mathcal H}\big(\bm{X}(\phi)|\mathcal H_m\big) = \prod_{i,n,\tau\in \mathcal{T}_\phi} p_{x|\omega}\big(x_{i,n}(\tau)|\hat{\omega}_{i,n}^m(\phi)\big)\;,
\label{pXH}
\end{equation}
where $m=0$ and $m=1$ denote the classification of {\em legitimate} and {\em fake}, respectively.

The details about the estimate of the parameters $\hat{\omega}^m_{i,n}(\phi)$ are provided in the next subsections. Plugging~\eqref{pXH} into \eqref{e:psi}, the LLR (\ref{e:psiphi}) is written as 
\begin{equation}
\Psi_\phi  = \sum_n R_{\phi, n}  
\label{e:psi2a}
\end{equation}
where 
\begin{align}
R_{\phi, n} &= \Psi_{\phi,n}^0 - \Psi_{\phi,n}^1\,,\label{e:psi2b} \\
\Psi_{\phi,n}^m &= \sum_{i,\,\tau\in \mathcal{T}_\phi}  \log  \frac{\hat{\beta}_{i,n}^m(\phi)}{2\hat{\sigma}_{i,n}^m(\phi)\Gamma\Big(\frac{1}{\hat{\beta}^m_{i,n}(\phi)}\Big)} \nonumber\\
& \quad\times {\rm e}^{-\left(\frac{|x_{i,n}(\tau)-\hat{\mu}^m_{i,n}(\phi)|}{\hat{\sigma}^m_{i,n}(\phi)}\right)^{\hat{\beta}^m_{i,n}(\phi)}}\,. 
\label{e:psi2}
\end{align}
From (\ref{e:psi2a})--(\ref{e:psi2}), we observe that the computation of the \ac{LLR} can be split among the $N$ trusted nodes: each trusted node estimates the parameter set $\hat{\omega}_{i,n}^m(\phi)$, computes the two \acp{LLR} for $m=0, 1$, and forwards $R_{\phi, n}$ to the sink. Then, at the sink, (\ref{e:psi2a}) is used to compute the packet LLR.

A bound for the system's performance, in terms of the FA and MD probabilities $P_{\rm FA}(\phi)$ and $P_{\rm MD}(\phi)$ for a given attack strategy can be obtained by applying the data processing inequality~\cite{7010914}. By defining the \ac{KL} divergence as
\begin{align}
{\mathbb D}(p_{\bm{X}|\mathcal H}&\big(\bm{X}(\phi)| \mathcal H_0)\,||\,p_{\bm{X}|\mathcal H}(\bm{X}(\phi)| \mathcal H_1)\big) \nonumber\\
&=\int p_{\bm{X}|\mathcal H}(\bm{X}| \mathcal H_0) \log\frac{p_{\bm{X}|\mathcal H}(\bm{X}| \mathcal H_0)}{p_{\bm{X}|\mathcal H}(\bm{X}| \mathcal H_1)} \, {\rm d}\bm{X}\,,
\label{defKL}
\end{align}
the bound is
\begin{equation}
f\!\big(\!P_{\rm MD}(\phi),\!P_{\rm FA}(\phi)\!\big) \!\leq\! {\mathbb D}\big(p_{\bm{X}|\mathcal H}(\bm{X}(\phi)| \mathcal H_0)||p_{\bm{X}|\mathcal H}(\bm{X}(\phi)| \mathcal H_1)\!\big),
\label{boundKL}
\end{equation}
where
\begin{align}
f\big(P_{\rm MD}(\phi),P_{\rm FA}(\phi)\big) &= P_{\rm MD}(\phi) \log \frac{P_{\rm MD}(\phi)}{1-P_{\rm FA}(\phi)} \nonumber\\
&\hspace{5mm}+\big(1-P_{\rm MD}(\phi)\big) \log \frac{1-P_{\rm MD}(\phi)}{P_{\rm FA}(\phi)}\,.
\label{boundKL2}
\end{align}
Under the assumption of independent errors, the \ac{KL} divergence can be written as
\begin{equation}
\begin{split}
&{\mathbb D}\big(p_{\bm{X}|\mathcal H}(\bm{X}(\phi)| \mathcal H_0)\,||\,p_{\bm{X}|\mathcal H}(\bm{X}(\phi)| \mathcal H_1)\big) \\
&=|\mathcal T_\phi| \sum_{i,n} {\mathbb D}\big(p_{x|\omega}(x_{i,n}(\tau)|\omega_{i,n}^0(\phi))\,||\,p_{x|\omega}(x_{i,n}(\tau)| \omega_{i,n}^1(\phi))\big)
\end{split}
\label{KLgen}
\end{equation}
The closed-form expression of the \ac{KL} divergence for two GG random variables is available~\cite{info1010013} for either $\beta_{i,n}^0(\phi) = \beta_{i,n}^1(\phi) = 2$, i.e., for  Gaussian distributions as
\begin{align}
{\mathbb D}&\big(p_{\bm{X}|\mathcal H}(\bm{X}(\phi)| \mathcal H_0)||p_{\bm{X}|\mathcal H}(\bm{X}(\phi)| \mathcal H_1)\big) \nonumber \\
&=|\mathcal T_\phi| \sum_{i,n} \log \frac{\sigma_{i,n}^1(\phi)}{\sigma_{i,n}^0(\phi)} - \frac{1}{2} + \left(\frac{\sigma_{i,n}^0(\phi)}{\sigma_{i,n}^1(\phi)} \right)^2 \nonumber \\
& \quad+ \frac{\big(\mu_{i,n}^1(\phi) - \mu_{i,n}^0(\phi)\big)^2}{2\big(\sigma_{i,n}^1(\phi)\big)^2}
\label{KL1}
\end{align}
or for $\mu_{i,n}^0(\phi) = \mu_{i,n}^1(\phi) = \mu$, a special case of the generalized Gamma distribution, where~\cite{Bauckhage-14}
\begin{align}
{\mathbb D} &\big(p_{\bm{X}|\mathcal H}(\bm{X}(\phi)| \mathcal H_0)\,||\,p_{\bm{X}|\mathcal H}(\bm{X}(\phi)| \mathcal H_1)\big) \nonumber\\
& = - \frac{1}{\beta_{i,n}^0(\phi)} + \log \frac{\beta_{i,n}^0(\phi) \sigma_{i,n}^1(\phi) \Gamma\left(\frac{1}{\beta_{i,n}^1(\phi)}\right)}{\beta_{i,n}^1(\phi) \sigma_{i,n}^0(\phi) \Gamma\left(\frac{1}{\beta_{i,n}^0(\phi)}\right)} \nonumber \\
& \quad + \frac{\Gamma\left(\frac{\beta_{i,n}^1(\phi)}{\beta_{i,n}^0(\phi)}\right)}{\Gamma\left(\frac{1}{\beta_{i,n}^0(\phi)}\right)}\left( \frac{\sigma_{i,n}^0(\phi)}{\sigma_{i,n}^1(\phi)}\right)^{\beta_{i,n}^1(\phi)} \,.
\label{KL2}
\end{align}
The computation of the \ac{KL} divergence for GG distribution with general parameters requires the numerical integration of~\eqref{defKL}.

\subsection{Estimation of the PDF Parameters at the Trusted Nodes}\label{sec:hypo}

In order to compute~\eqref{e:psi2a} and~\eqref{e:psi2b}, each trusted node $n$ must estimate the set of \ac{PDF} parameters, $\hat{\omega}^m_{i,n}(\phi)$, for the two hypotheses $m=0$ and $m=1$. Note that 
the procedure to classify the packets as authentic or not authentic should not imply feedback from the sink to the trusted nodes, as this would increase the authentication overhead, would also have to be secured against, e.g., forging or jamming attacks on the feedback packet. For this reason, no consensus procedures can be implemented. 
Instead, here we do not employ any feedback, but rather propose that each trusted node estimates all parameters and associates them with the two hypotheses.

\noindent \emph{Parameter Estimation} --- As the estimation criterion, we consider the maximum likelihood approach, wherein parameters are selected in order to maximize the probability of the observations. Given the parameters, we employ the mixture model
\begin{align}
{\mathbb P}\big({\cal X}_{i,n}(\phi)|&\hat{\omega}_{i,n}^0(\phi),\hat{\omega}_{i,n}^1(\phi)\big)  \nonumber\\
&= \prod_{t\in \mathcal{T}_f} \sum_{m=0}^{1}k^m_{i,n} \, p_{x| \omega}\big(x_{i,n}(t)|\hat{\omega}^m_{i,n}(\phi)\big)\;,
\end{align}
where ${\cal X}_{i,n}(\phi) =\{x_{i,n}(t), t\in {\cal T}_{f}, f \leq \phi\}$ (i.e., we use all observations collected so far), and $k^m_{i,n}$ is the prior on parameter $i$ at node $n$ about hypothesis $m$ (which is also unknown and must be estimated). We let each trusted node estimate the set $\theta_{i,n}^m = (\omega_{i,n}^m, k_{i,n}^m)$.

Let us now define the set $\Theta_{i,n} = \{\theta_{i,n}^m, m=0,1\}$ and similarly the corresponding set of estimated parameters at packet $\phi$ as  $\hat{\Theta}_{i,n}(\phi)$. Note that $\theta_{i,n}^m$ is not packet-dependent, in accordance with our assumption that the feature statistics for both the legitimate node and the attacker do not change significantly across different packets. However, the estimate $\hat{\theta}_{i,n}^m(\phi)$ is updated after each packet. The estimated parameters $\hat{\Theta}_{i,n}(\phi)$ maximize the likelihood of observing $\mathcal X_{i,n}(\phi)$. Taking into account the fact that the observations $x_{i,n}(t)$ are independent for different $t$ (even when belonging to the same packet), we have
\begin{equation}
\hat{\Theta}_{i,n}(\phi) = \argmax_{\,\bar{\Theta} = (\omega_{i,n}^m,k_{i,n}^m)} {\mathbb P}({\cal X}_{i,n}(\phi)|\Theta_{i,n}=\bar{\Theta})\;.
\label{e:model}
\end{equation}
We find the solution of the maximization problem \eqref{e:model} via the expectation maximization (EM) algorithm~\cite{Mudd_EM}, as detailed in the Appendix.

\noindent \emph{Parameter Association} --- The EM algorithm provides the estimated parameter sets $\hat{\theta}_{i,n}^m(\phi)$, for $m=0$ and $m=1$. However, without further information about the expected values for each of the two classes $m=\{0,1\}$, the association of $\hat{\theta}^m_{i,n}(\phi)$ to hypothesis ${\cal H}_0$ or ${\cal H}_1$ is ambiguous. In other words, we do not know if $m=0$ is actually associated with hypothesis $\mathcal H_0$ or hypothesis $\mathcal H_1$. Therefore, when fusing all data at the sink, we must synchronize $m$ across the trusted nodes to ensure that the same value of $m$ refers to the same hypothesis for all the trusted nodes. 

According to our assumption that the first packet always originates from the legitimate transmitter, all trusted nodes assign hypothesis $m=0$ to the first packet with index $\phi=0$. Then, the labeling $m$ is chosen by matching the current estimated variables $\hat{\theta}^m_{i,n}(\phi)$ to the measurements from the first packet, ${\cal X}_{i,n}(0)$. Formally, we associate $\hat{\theta}^0_{i,n}(\phi)$ with hypothesis ${\cal H}_0$ if
\begin{equation}
 {\mathbb P}\big(\mathcal X_{i,n}(0)|\hat{\theta}_{i,n}^0(\phi)\big) \geq {\mathbb P}\big(\mathcal X_{i,n}\big(0)|\hat{\theta}_{i,n}^1(\phi)\big)\,,
\label{e:condH0}
\end{equation}
and from (\ref{e:distribution}), condition (\ref{e:condH0}) becomes 
\begin{equation}
\prod_{t \in \mathcal T_0} p_{x|\omega}\big(x_{i,n}(t)|\hat{\omega}_{i,n}^0(\phi)\big) \geq \prod_{t \in \mathcal T_0} p_{x|\omega}\big(x_{i,n}(t)|\hat{\omega}_{i,n}^1(\phi)\big)\,.
\label{e:assosiate1}
\end{equation}
After this association procedure, for $m=0$, trusted node $n$ calculates~\eqref{e:psi2} and then the decision index $R_{\phi,n}$ in~(10), according to the $\hat{\theta}_{i,n}^m(\phi)$ tuples associated with hypotheses ${\cal H}_0$, and 
${\cal H}_1$. 

\subsection{Refined Data Fusion} \label{sec:refdatafusion}

We observe that in the above distributed procedure, two possible errors may occur: a) a hypothesis association error, and b) a parameter estimation error. 
The first error occurs when $m$ is assigned different values for the same hypothesis by different nodes, i.e., when test~\eqref{e:condH0} fails. The probability of occurrence of this error $P_e$ can not be obtained in closed form. However, when the number of samples goes to infinity (i.e., the products in (\ref{e:assosiate1}) have a large number of terms) we can resort to the Chernoff-Stein lemma \cite[Theorem 11.8.3]{Cover}, which states that the error probability goes exponentially to zero as a function of  $|\mathcal T_0|$, with exponent given by the \ac{KL} divergence  (\ref{KLgen}) computed for $\phi =0$, i.e.,
\begin{equation}
\lim_{|\mathcal T_0| \rightarrow \infty} \frac{1}{|\mathcal T_0|} \log P_e \!=\! -{\mathbb D}\big(p_{\bm{X}|\mathcal H}(\!\bm{X}(0)| \mathcal H_0)||p_{\bm{X}|\mathcal H}(\!\bm{X}(0)| \mathcal H_1)\!\big),
\label{e:bound1}
\end{equation}
provided that the parameters $\hat{\omega}^m_{i,n}$ have been correctly estimated.

To address the hypothesis association error, since the sink node collects more information than each single trusted node, we let the sink correct the association by flipping the sign of the reported $R_{\phi, n}$ as follows. The parameter estimation error occurs when the observed data is noisy and does not represent the full feature statistics, or when the attacker manages to confuse a trusted node. The sink node can still make a robust decision by weighing the values $R_{\phi,n}$. The two refinements are jointly obtained by replacing~\eqref{e:psi2a} with 
\begin{equation}
\Psi_{\phi} = \sum_{n} g_n(\phi) \, s_n(\phi) \, R_{\phi,n}\;,
\label{e:decision}
\end{equation}
where $g_n(\phi) >0$ is a weight to correct PDF parameter estimation errors, and $s_n(\phi) \in \{-1, 1\}$ is the binary correction factor for the hypothesis association.

\noindent \emph{Choice of $g_n(\phi)$} --- The weighting mechanism reflects the fact that the physical location of the trusted node matters in two respects: if a trusted node $n$ is closer to the packet source, be it the legitimate transmitter or the attacker, the estimate $\hat{\omega}^m_{i,n}$ is more reliable. Moreover, if the trusted nodes are spread around the network area, instead of being clustered close to one another, it becomes easier to detect that the attacker is imitating a certain transmitter-trusted node channel.
This is because for each fake packet sent, the attacker can realistically attempt to mimic the channel only within a limited area.

We use channel feature $i^\star=6$ (the smoothed received power in Section~\ref{sec:param}) as a measure of physical proximity to the packet source,
and formalize the weighing function as
\begin{equation}
g_n(\phi)=\Bigg(\frac{\bar{x}_{i^\star,n}(\phi)}{\underset{n'}{\max}(\bar{x}_{i^\star,n'}(\phi))}\Bigg)^{-2}(1+\zeta_{n})\;,
\label{e:weight}
\end{equation}
where $\bar{x}_{i^\star,n}(\phi)$ is the mean of the set $\{x_{i^\star,n}(t) , t \in \mathcal{T}_\phi\}$. 
The first term of \eqref{e:weight} is the normalized weighting function, while $\zeta_{n}$ ensures that isolated trusted nodes are given higher weights than clustered nodes, as they contribute with more information. To that end, we choose $\zeta_{n}$ to reflect the variance of the UTM locations of the trusted nodes, normalized to the closeness to the packet source. Specifically, $\zeta_{n}=\sum_{k\neq n}\zeta_{n,k}$,
where
\begin{align}
\zeta_{n,k}&=\frac{1}{\sum_{j\neq k,n}\bar{x}_{i^\star,j}(\phi)} \cdot \sum_{j\neq k,n}\frac{1}{\bar{x}_{i^\star,j}(\phi)} \nonumber \\
&\qquad\times \left(1-\frac{\big(\ve{p}(k)-\ve{p}(n)\big)^{\mathrm{T}}\big(\ve{p}(j)-\ve{p}(n)\big)}{\bar{x}_{i^\star,n}(\phi)\bar{x}_{i^\star,k}(\phi)}\right)
\label{e:weight2}
\end{align}
is a normalized distribution measure between nodes $n$ and $k$, and $\ve{p}(\ell)$ is the set of UTM coordinates of node $\ell$.

\noindent \emph{Choice of $s_n(\phi)$} --- We choose $s_n(\phi)$ at the sink node so as to maximize the agreement among the trusted nodes. We adopt a majority rule, i.e., let $\bar{s}_{\phi} \in \{-1, 1\}$ be the sign of the majority of reports $R_{\phi,n}$, then
\begin{equation}
s_n(\phi) = {\rm sgn}(R_{\phi,n}) \bar{s}_{\phi},
\label{e:sn}
\end{equation}
so that ${\rm sgn}(s_n(\phi) R_{\phi,n}) =\bar{s}_{\phi}$ for all $n$. Also in this case an analysis of the probability of error in the sign change does not yield closed-form expressions. Assuming that the choice of the majority sign $\bar{s}_\phi$ is correct (i.e., it corresponds to the true hypothesis), and that both the parameters and the assignment~\eqref{e:condH0} are correct, we consider the probability that any of the $N$ signs of $R_{\phi,n}$ is not correct. Let $\pi_n = \mathbb P[{\rm sgn}(R_{\phi, n}) \neq \bar{s}_\phi]$. In this case, since the parameters and the assignment are correct, we should not change the sign, and the probability that this error occurs for at least one $n$ is
\begin{equation}
P_{e, s} = 1 - \prod_n (1 - \pi_n)\,.
\label{e:sn2}
\end{equation}
Now, for small values of $\pi_n$ we have $P_{e, s} \approx \max \pi_n$. Therefore, resorting again to the Chernoff-Stein lemma, we can conclude that $P_{e, s}$ goes to zero as $|\mathcal T_\phi| \rightarrow \infty$ exponentially with exponent
\begin{align}
\lim_{|\mathcal T_\phi| \rightarrow \infty} & \frac{1}{|\mathcal T_\phi|} \log P_{e, s} \nonumber \\
& \quad\approx - \min_n |\mathcal T_\phi| \sum_{i } {\mathbb D}\big(p_{x|\omega}(x_{i,n}(\tau)|\omega_{i,n}^0(\phi))\,|| \nonumber \\
& \qquad ||\,p_{x|\omega}(x_{i,n}(\tau)| \omega_{i,n}^1(\phi))\big),
\label{e:sn3}
\end{align}
where the approximation comes from having considered only the error probability of the least reliable node $n$, i.e., the one having the minimum \ac{KL} divergence.

Algorithm~1 summarizes the proposed authentication procedure. The trusted nodes act first (lines~3--7), and the sink makes the final decision (lines~9--16). 
The complexity of the EM algorithm is ${\cal O}\left(N_{\mathrm{sym}}N_{\rm EM}\right)$~\cite{Mudd_EM}, where $N_{\mathrm{sym}}$ is the number of elements in ${\mathcal X}(\phi)$ and $N_{\rm EM}$ is the number of EM iterations. The numerical solution to \eqref{e:sol_a}--\eqref{e:sol_c} is obtained through the alternating optimization approach~\cite{wang2015alternating}, whose complexity is ${\cal O}\left(N_{\mathrm{sym}}^3 + N_{\mathrm{sym}}^2 + N_{\mathrm{sym}}\right)$. The EM procedure is performed for each of the four estimated channel feature, and for each trusted node. Finally, the authentication decision has complexity ${\cal O}\left(1\right)$. Thus, the complexity of our algorithm for authenticating a single packet is ${\cal O}\left(4N(N_{\mathrm{sym}}+N_{\mathrm{sym}}^3 + N_{\mathrm{sym}}^2)N_{\rm EM}\right)$. We also note that the EM algorithm, as well as the alternate maximization process, provably converge to a local maximum of the log-likelihood function (\ref{e:log}). We explore this convergence numerically in the following.

\begin{algorithm}[t]
\caption{Cooperative Authentication Algorithm} \label{alg1}
\begin{algorithmic}[1]
\REQUIRE Set of channel measurements $\{x_{i,n}(t) , t \in \mathcal{T}_\phi\}$, $i \in {\cal I}$, for each trusted node $n$ and for each current and previous packets $\phi$; \ENSURE Authentication decision for a packet $\phi$;
\renewcommand{\algorithmicensure}{\textbf{Begin}} \ENSURE  
\STATE Calculate $\omega_{i,n}^m, \ m=0,1$ (see Section V); 
\FOR{each received packet $\phi$}
\FOR{each {\bf trusted node} $n$}
\STATE Associate hypothesis $m=\{0,1\}$ by \eqref{e:assosiate1};
\STATE Calculate $\Psi_{\phi,n}^m$ from \eqref{e:psi2};
\STATE Calculate local decision index $R_{\phi, n}$ by \eqref{e:psi2b};
\STATE Forward $R_{\phi, n}$ to sink;
\ENDFOR
\STATEx {\bf{\hspace{0.2cm} At the sink:}}
\STATE Calculate hypothesis correction $s_n(\phi)$ by \eqref{e:sn};
\STATE Calculate trusted nodes' weights $g_n(\phi)$ by \eqref{e:weight};
\STATE Calculate decision index $\Psi_{\phi}$ by \eqref{e:decision};
\IF{$\Psi_{\phi} \geq \lambda$} \STATE packet is authentic; \ELSE \STATE packet is false;
\ENDIF
\ENDFOR
\renewcommand{\algorithmicensure}{\textbf{End}} \ENSURE
\end{algorithmic}
\end{algorithm}

\section{Numerical Results}\label{sec:numres}

\begin{figure}[t]
       \centering
       \includegraphics[width=8cm]{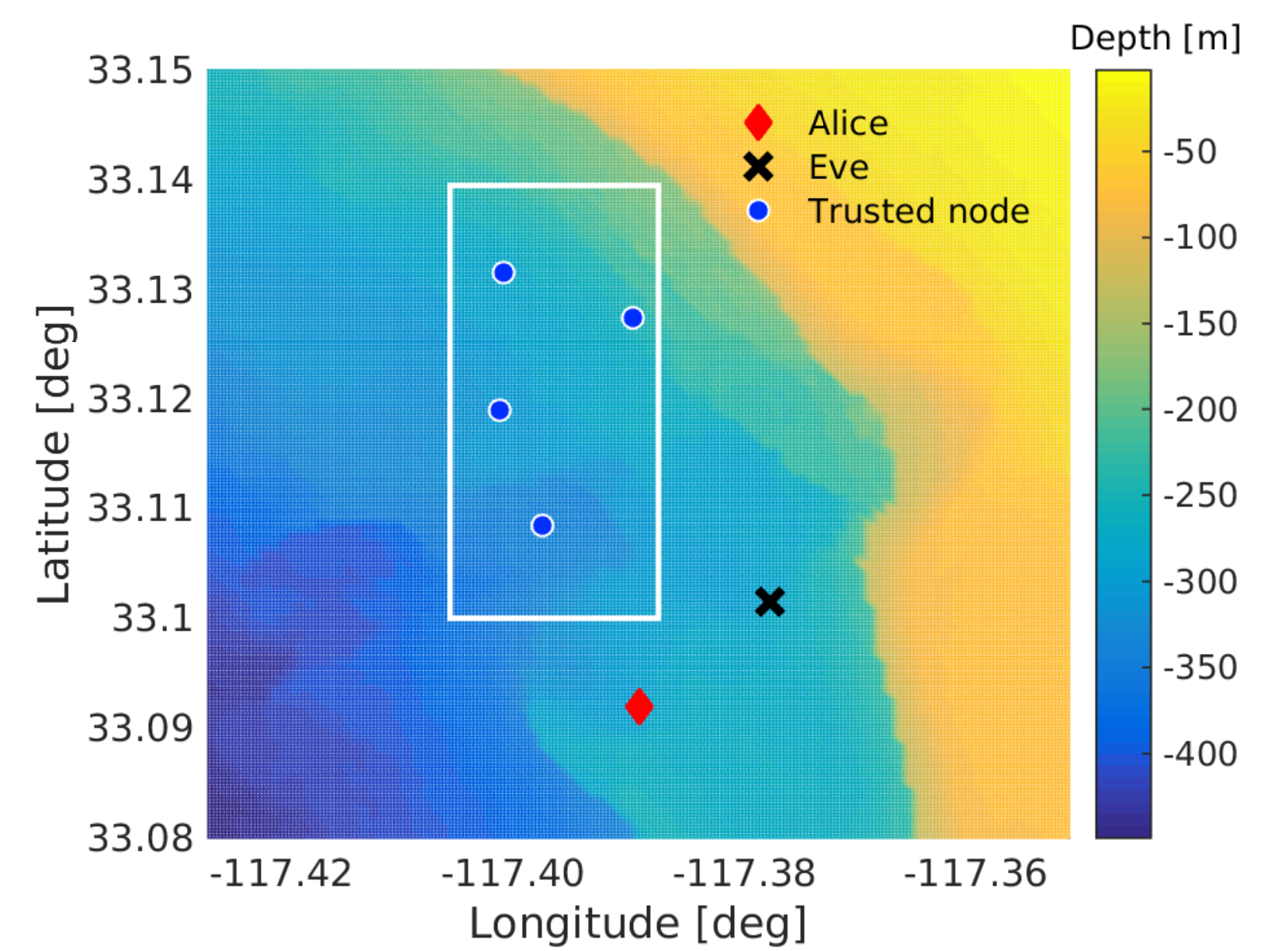}
       \caption{Bathymetry map of the area, showing the location of Alice and the trusted node deployment area.}
       \label{f:example_bhop_map}
\end{figure}

We evaluate our authentication approach using realistic channel simulations obtained through the Bellhop framework~\cite{bellhop}, an established ray-tracing tool to simulate acoustic propagation under water. We set our simulations in the Bay Area of San Diego, CA, between the latitude/longitude coordinates $[32.6^\circ,-117.8^\circ]$ and $[33.3^\circ,-117.2^\circ]$. The setting is characterized by a continental shelf zone with a quasi-flat bottom, of a depth between 50 and 100~m, followed by a ridge, which plunges to lower depths. Here we consider a portion of this area, whose bathymetry map is shown in Fig.~\ref{f:example_bhop_map}. In this portion, the depth ranges from about 10~m in the top-right corner of the map, to about 450~m in the bottom-left corner. We assume that the bottom sediments are mostly sandy and that the sound speed profile decreases almost linearly from 1520~m/s (sea level) to 1480~m/s (depth 450~m),
which corresponds to an actual measurement taken in the area during the summer season. Furthermore, we assume a flat ocean surface.

Fig.~\ref{f:example_bhop_map} also provides a view of our simulation setup. The locations of the legitimate transmitter (Alice) and the attacker (Eve) are shown as a red diamond and a black cross, respectively. The distance between the legitimate node and the attacker is about 1500~m. 
The trusted nodes are deployed within the area enclosed in the white frame, which has a size of about 2~km $\times$ 2~km (notice that the scale is different on the x-axis and y-axis in Fig.~\ref{f:example_bhop_map}). 
As the focus of this evaluation is on the authentication model itself and not on the design of the communication between the trusted node and the sink, we do not explicitly model the links between the sink and the trusted nodes in the following.

We collect a Monte-Carlo set of 500 simulation runs. In each run, the location and depth of the trusted nodes are uniformly drawn at random, whereas the locations of the legitimate node and the attacker remain fixed. Each simulation run corresponds to the transmission of a total of six packets from the legitimate transmitter to all trusted nodes, and of one packet from the attacker, also directed to all trusted nodes.
Each packet includes 100 symbols, out of which each trusted node produces 100 estimates of the channels' number of taps \eqref{e:TapNum}, relative RMS delay spread \eqref{e:Delay}, and the smoothed received power \eqref{e:smooth}, which have been found to offer sufficient diversity to enable our authentication scheme. Since the transmitted signals were wideband (relative to the carrier frequency) and transmitted with high SNR, channel estimates were obtained by processing the received signals through a normalized matched filter. Relevant channel taps were determined by identifying the peaks exceeding a threshold corresponding to a probability of false alarm of $10^{-4}$ according to the analysis of~\cite{MF}. For each packet, we obtain a different channel estimate by simulating the oscillation of the nodes around their mooring location. For each symbol, we perform a separate Bellhop run where the transmitter and the receiver are randomly displaced by a short distance from their nominal location. Assuming a realistic transmission system with 10~kHz of bandwidth, the power-delay profiles have been filtered such that no two arrivals exist whose delays are less than 100~$\mu$s apart. The correlation among different channel realizations is ensured by the fact that the displacement of each node is small.

We compare our scheme with a typical approach for authentication that is also aligned with our proposal, referred to as \textit{benchmark}. This approach compares the 2-norm difference between a feature of a given impulse response and the same feature of a reference impulse response. Following~\cite{hao13,xiao15}, we choose the received power as the compared channel feature, and the reference channel to be that of the first packet of the legitimate transmitter.

\subsection{Attack Strategy} \label{sec:attstrat}

To implement the attacker model in Section~\ref{sec:attackermodel}, we consider different attacker capabilities which, in turn, enable increasingly complex attack strategies. The most typical case is a straightforward impersonation attack, where the attacker tries to disguise himself as the legitimate transmitter by sending packets to all trusted nodes. This {\em na\"{\i}ve} attack can succeed if, e.g., the attacker is close to the legitimate node. We model this attack by using the Bellhop software~\cite{bellhop} to estimate the channel impulse response from the legitimate node or the attacker to all of the trusted nodes. 

A more powerful attack, referred to as \textit{TN-1}, is enabled if the attacker can estimate the channel impulse response between the legitimate node and one of the trusted nodes, and can leverage an array of transducers in order to approach this channel response. To compute the channel estimate, we assume the attacker knows the exact sound speed profile and the location of both the legitimate and the trusted nodes, but has an imperfect bathymetry map. Specifically, the attacker stores the value of the bathymetry for the same set of latitude and longitude coordinates used in the simulation, but each bathymetry sample is affected by a constant offset $\Delta z$ equal to 5\% of the maximum depth, and by an error drawn at random in the interval $[-0.1\Delta z,0.1\Delta z]$. The attacker runs the same acoustic propagation model provided by Bellhop to estimate the channel impulse response and then pre-processes its transmission, such that it will be received as having gone through the estimated channel between the legitimate node and the trusted node.

An even more complex attack is obtained by assuming that the attacker can approximate the channel response between the legitimate node and multiple trusted nodes, ultimately all of them. This attack is referred to as \textit{TN-$x$}, and makes it possible for the attacker to disguise itself as the legitimate node, limited only by its capability to estimate the channels between the legitimate node and the trusted nodes.

We stress that trying to fool a single trusted node in TN-1 requires several complex steps, including channel estimation, channel inversion and pre-coding, but may still be possible for an underwater attacker endowed with a large transceiver array and sufficient computational power. Instead, fooling multiple cluster nodes at the same time in TN-$x$, $x>1$, would require the deployment of multiple, perfectly coordinated transceiver array elements at different locations in the network area. This is orders of magnitude less realistic in any practical underwater scenario.

\subsection{Determining the Threshold level}\label{sec:th}

In order to make the authentication decision in \eqref{e:threshold}, we need to determine the decision threshold $\lambda$. We offer two approaches. The first is to determine the receiver operating characteristic (ROC) curve as a function of the threshold (\textit{ROC test}), either theoretically using the relation in \ref{boundKL} or numerically. The second method is based on a support vector machine (SVM) to classify the two hypotheses based on the decision index ({\em SVM-test}). For both methods, we divide the simulation data into a \textit{training} part and a \textit{test} part. The threshold computation is performed only on the training data set, whereas the data analysis is performed only on the test set. 

In the ROC test, we try a set of threshold values over the training data and draw the TP rate against the FN rate. We then determine a desirable working point that trades off between the values of these two metrics. In the SVM test, we avoid setting such a trade-off, and instead seek the threshold that yields the best classification solution. To this end, we perform a $K$-fold test to determine the classifier's parameters, and train the SVM model based on the training set. Classification is performed on the test set, and the threshold is determined based on the minimum decision index from all packets classified as belonging to the attacker.

\subsection{Simulation Results}\label{sec:simres}

\begin{figure}
\centering
\includegraphics[width=8cm]{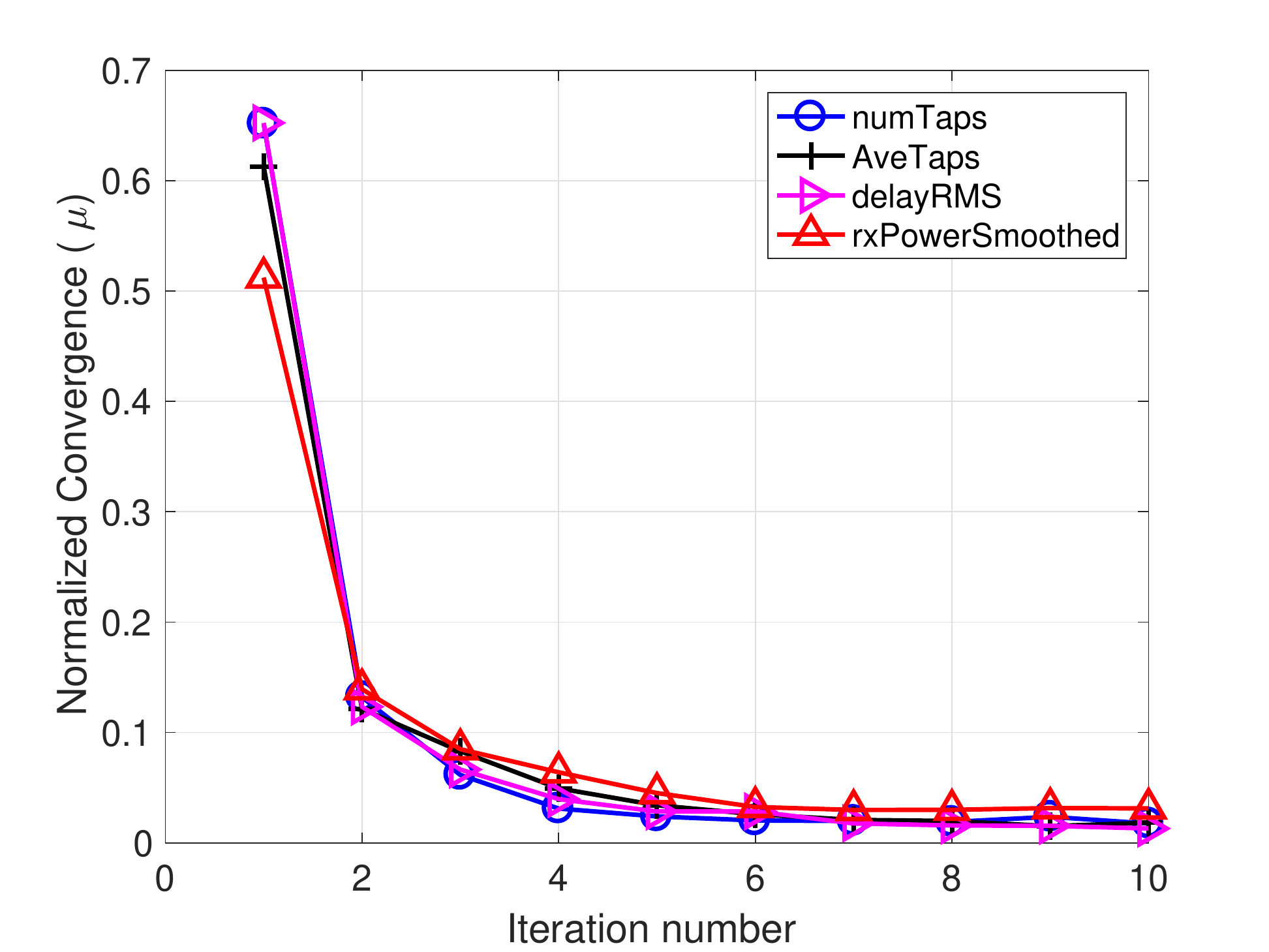}
\caption{Average convergence of the EM algorithm for parameter $\mu$.}
\label{f:Convergence}
\end{figure}

We first describe the simulation results for the {\em na\"{\i}ve} attacker, which transmits its packets directly, without any preprocessing. Fig.~\ref{f:Convergence} shows the normalized convergence of the EM algorithm for the mean parameter $\mu$, averaged over the trusted nodes and the number of simulations for all estimated parameters. We observe that, for all estimated parameters, the EM procedure converges after only 6 iterations on average. Similar results are obtained for the convergence of $\sigma$ and $\beta$.

\begin{figure}
\centering
  \subfloat[ROC curve for the KL bound~\eqref{boundKL}, GG signals with estimated parameters (GG Model), and Bellhop signals (Actual).]{\label{f:ROCSimulationnaive}\vspace{0.1in}\includegraphics[width=3.2in]{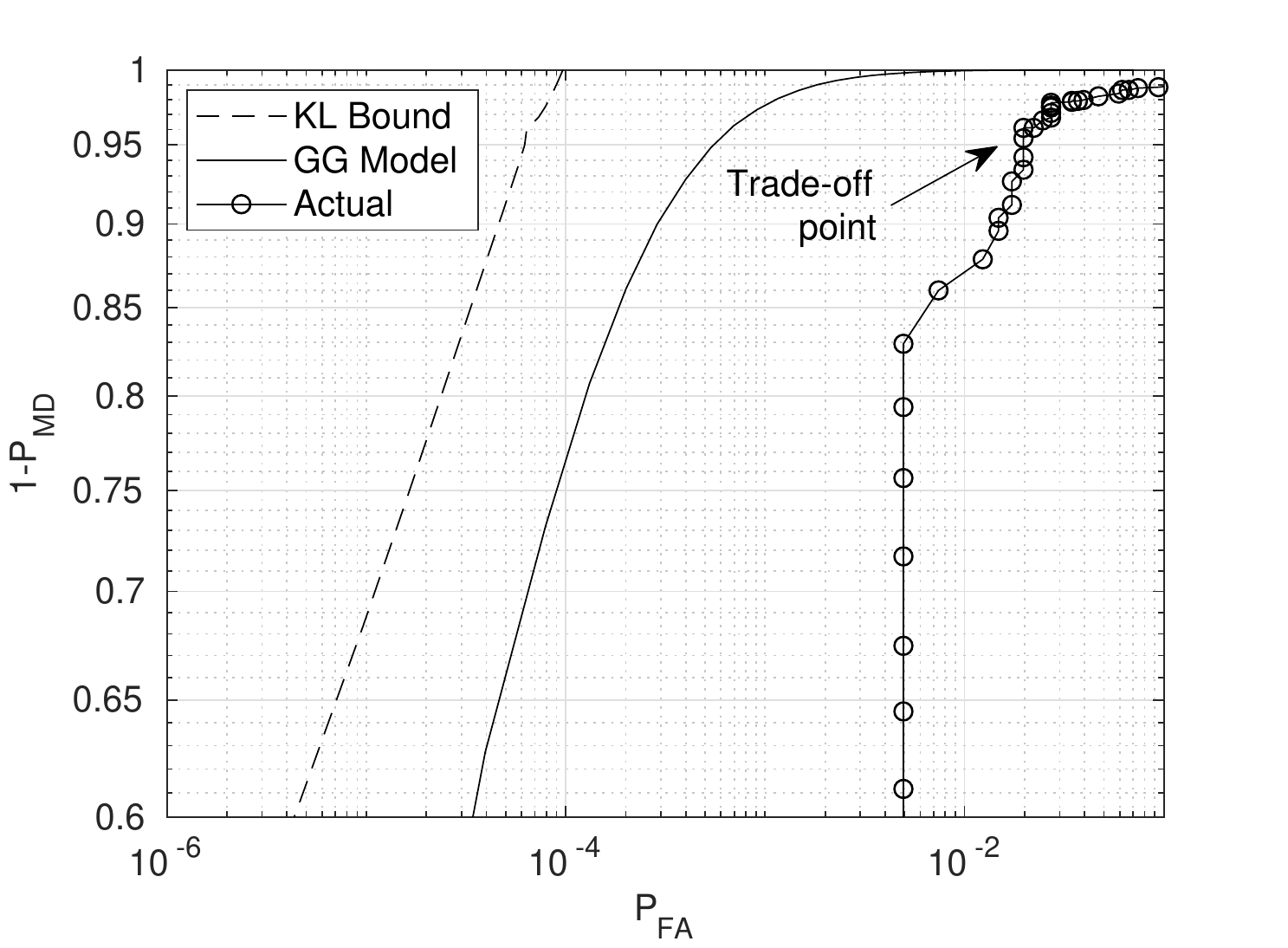}}%
  \hfill
  \subfloat[Empirical distribution of the decision index for our method and the benchmark.]{
  \label{f:HistSimulationnaive}\includegraphics[width=3.1in]{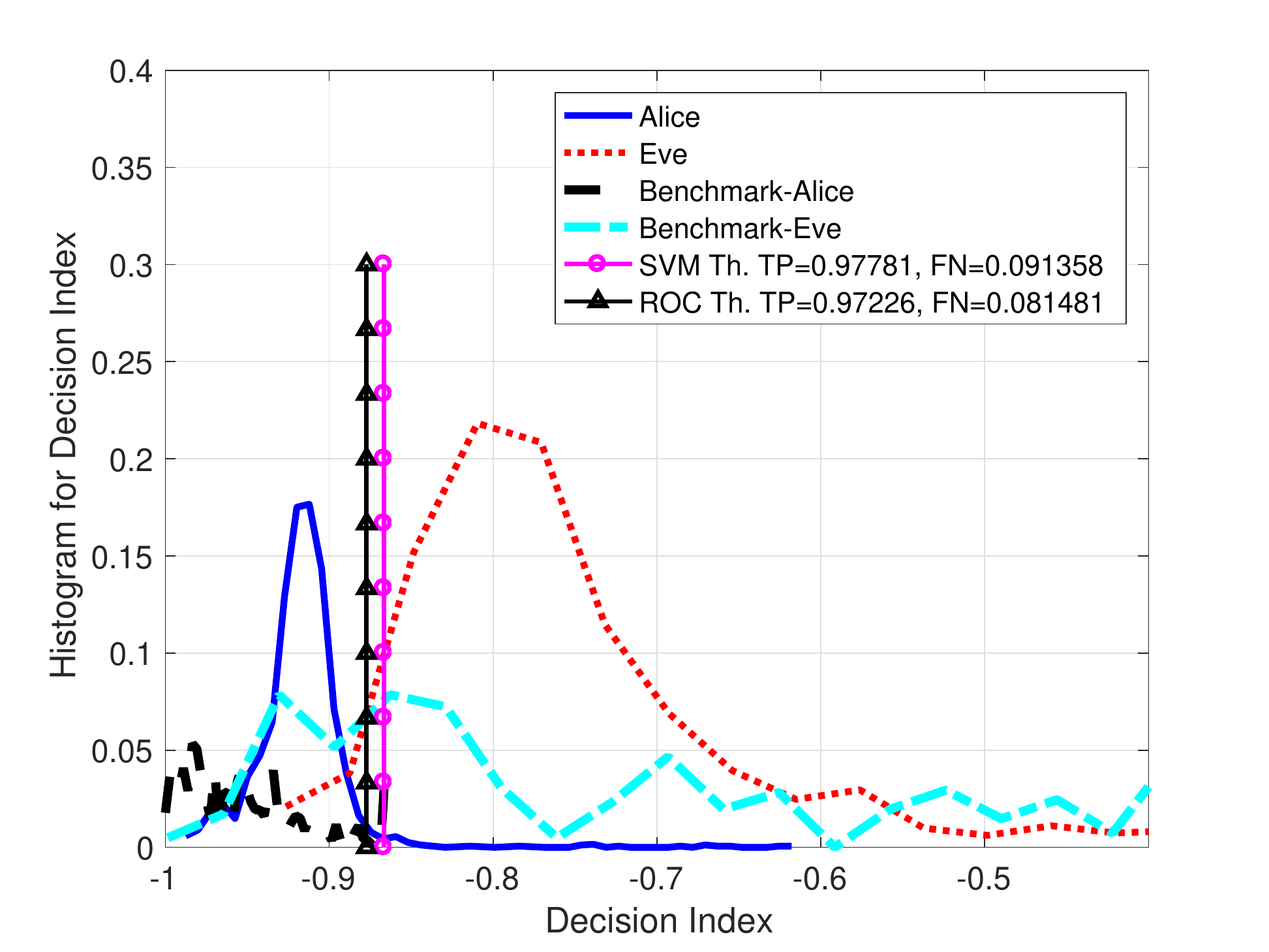}}
  \caption{Simulation results of LLR $\Psi_{\phi}$ from \eqref{e:threshold} for a na\"{\i}ve attacker. The obtained authentication accuracy is $97\%$.} 
\label{f:Simnaive}
\end{figure}

In Fig.~\ref{f:ROCSimulationnaive} we assess the average ROC from a) bound~\eqref{boundKL} on GG signals with  estimated parameters (KL Bound), b) LLR thresholding on GG signals with estimated parameters (GG Model), c) LLR thresholding on Bellhop simulations (Actual). In particular, for a) and b), 
the bound and ROC are obtained using signals generated according to the GG model and the estimated parameters, respectively. For c) we apply the detection technique described in the previous subsection for the estimated parameters from Bellhop simulations. For all scenarios, we obtain a ROC curve for each parameter estimation, then for each misdetection probability we average the false alarm probability to obtain average ROC and bound curves. As expected, the bound provides the highest true positive ($1 - P_{\rm MD}$) probability for the same false alarm probability, while the ROC obtained from the GG model has a slightly reduced true positive probability. The effective ROC shows worse performance due to estimation errors affecting the GG parameters and to a slight model mismatch between the Bellhop simulations and the GG model.

Fig.~\ref{f:HistSimulationnaive} shows the histogram of the decision index, normalized within the $[-1,1]$ range, for 1,000 simulated test scenarios, together with the performance of the benchmark scheme, the SVM test-based threshold, and the ROC test-based threshold. The latter was set based on the ROC curve in Fig.~\ref{f:ROCSimulationnaive}, and we choose as a best practice (at least under the assumption that the channel does not vary significantly) the {\em knee} point in the ROC curve to yield a desired FN probability of 0.1 and a TP probability of 0.98. The SVM test-based threshold was obtained from a radial basis function (RBF) kernel SVM with a $K=5$-fold training procedure. The procedure achieved an excellent authentication accuracy of $97\%$. We observe that the decision index values for the legitimate packets (Alice) and the attacker's packets (Eve) are well distinguished. While a similar trend is also obtained for the benchmark scheme, here the difference between Alice and Eve is less dominant. Although they have been obtained through different procedures, the threshold based on the SVM test is close to the one based on the ROC test. However, the SVM-based threshold yields a slightly better performance (the same TP, but lower FN probability).

\begin{figure*}[t]
\centering
  \subfloat[TN-1, Accuracy 96\%.]{\label{f:HistSimSmartEveNode1}\includegraphics[width=2.8in]{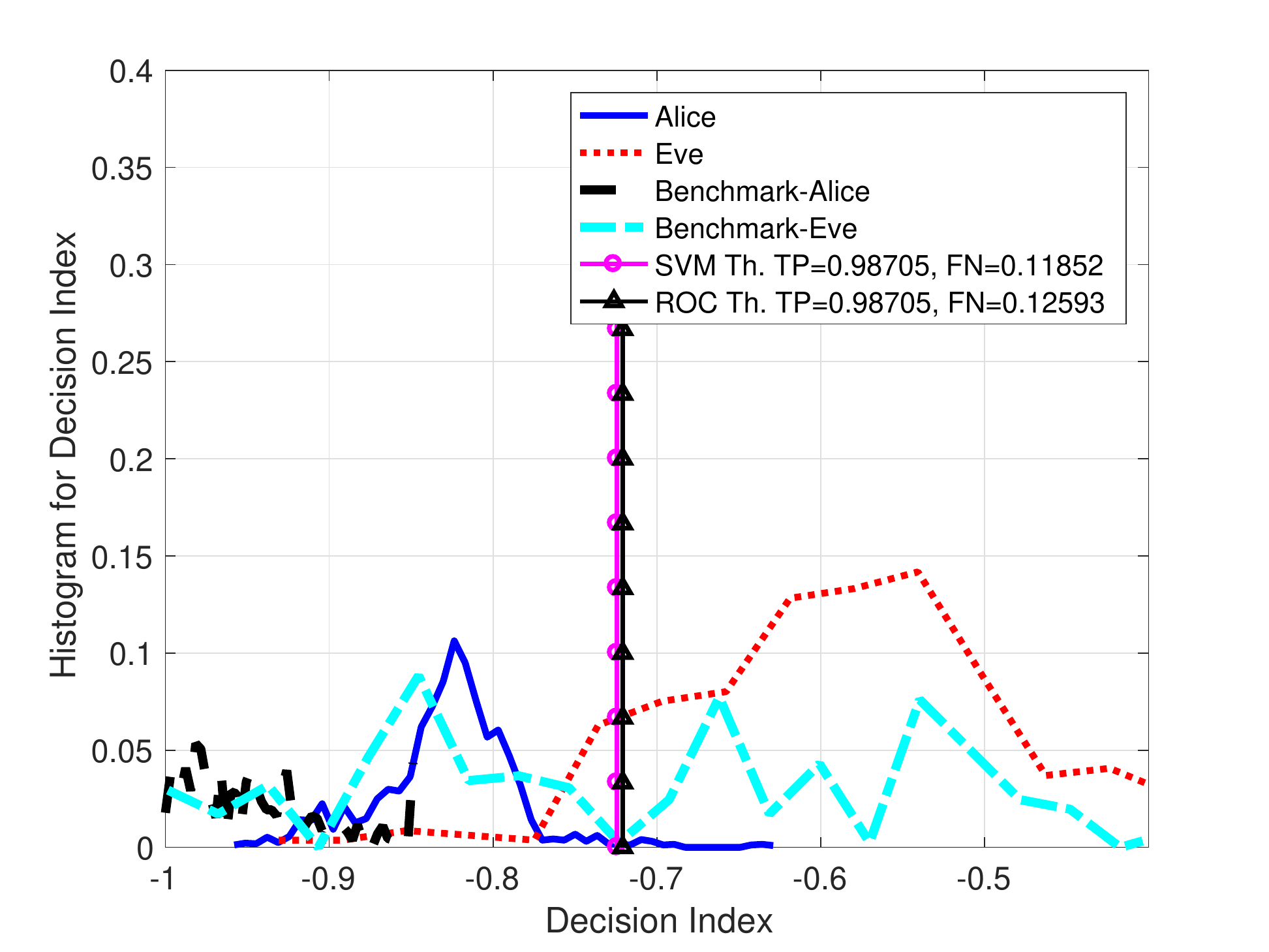}}\hspace{1em}%
  \subfloat[TN-2, Accuracy 96\%.]{\label{f:HistSimSmartEveNode2}\includegraphics[width=2.8in]{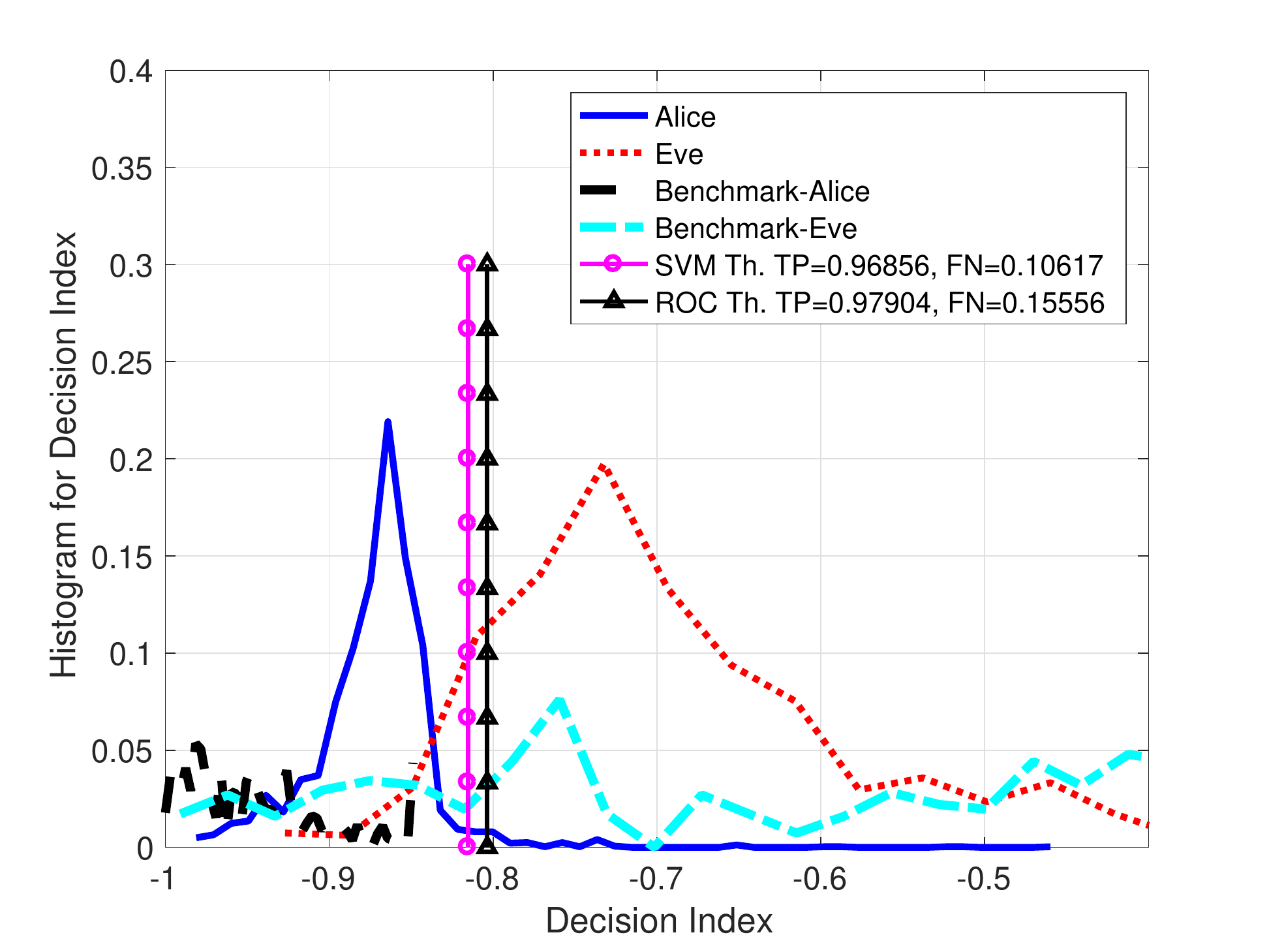}}\\[-1mm]
  \subfloat[TN-3, Accuracy 94\%.]{\label{f:HistSimSmartEveNode3}\includegraphics[width=2.8in]{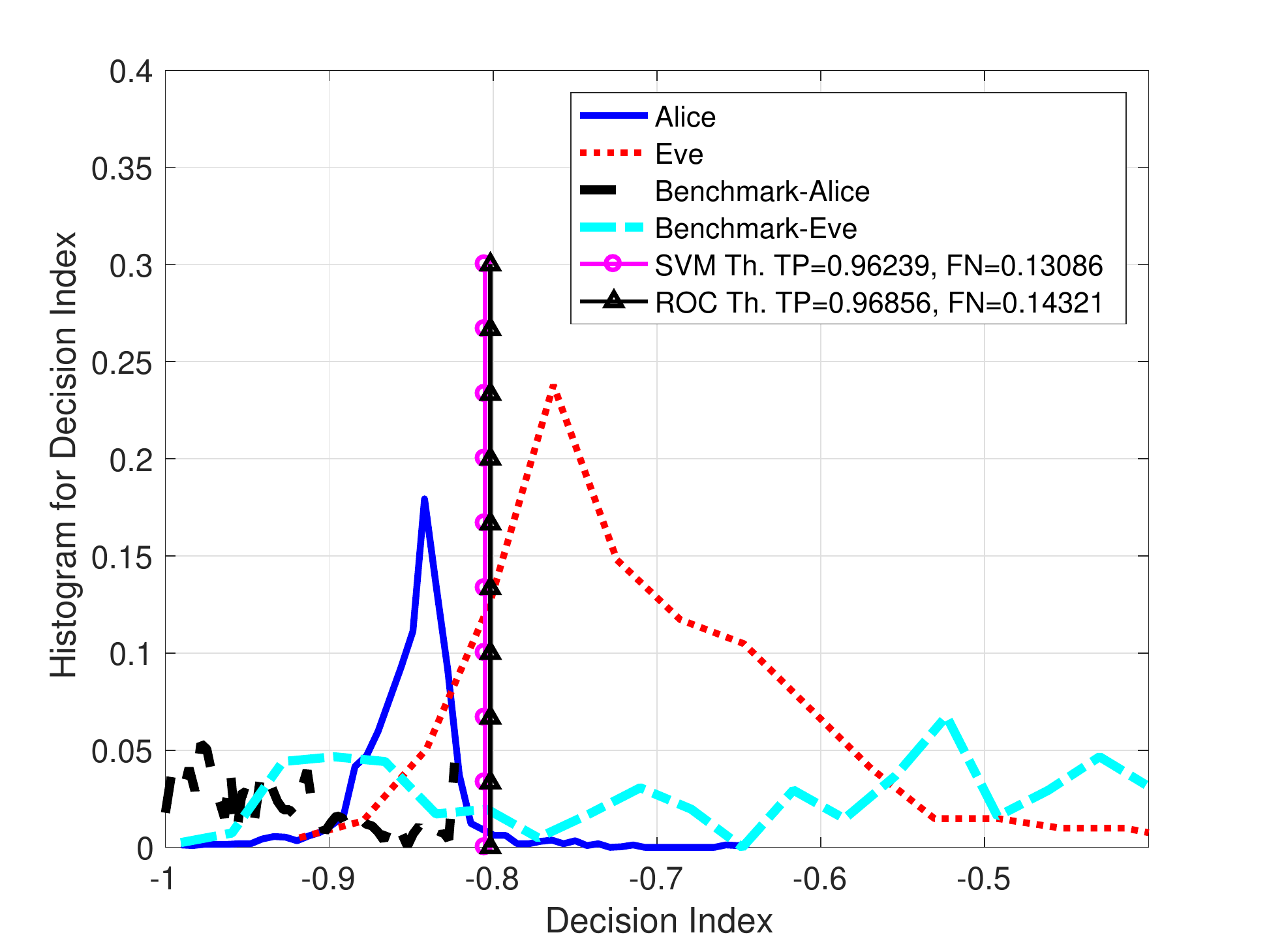}}\hspace{1em}%
  \subfloat[TN-4, Accuracy 92\%.]{\label{f:HistSimSmartEveNode4}\includegraphics[width=2.8in]{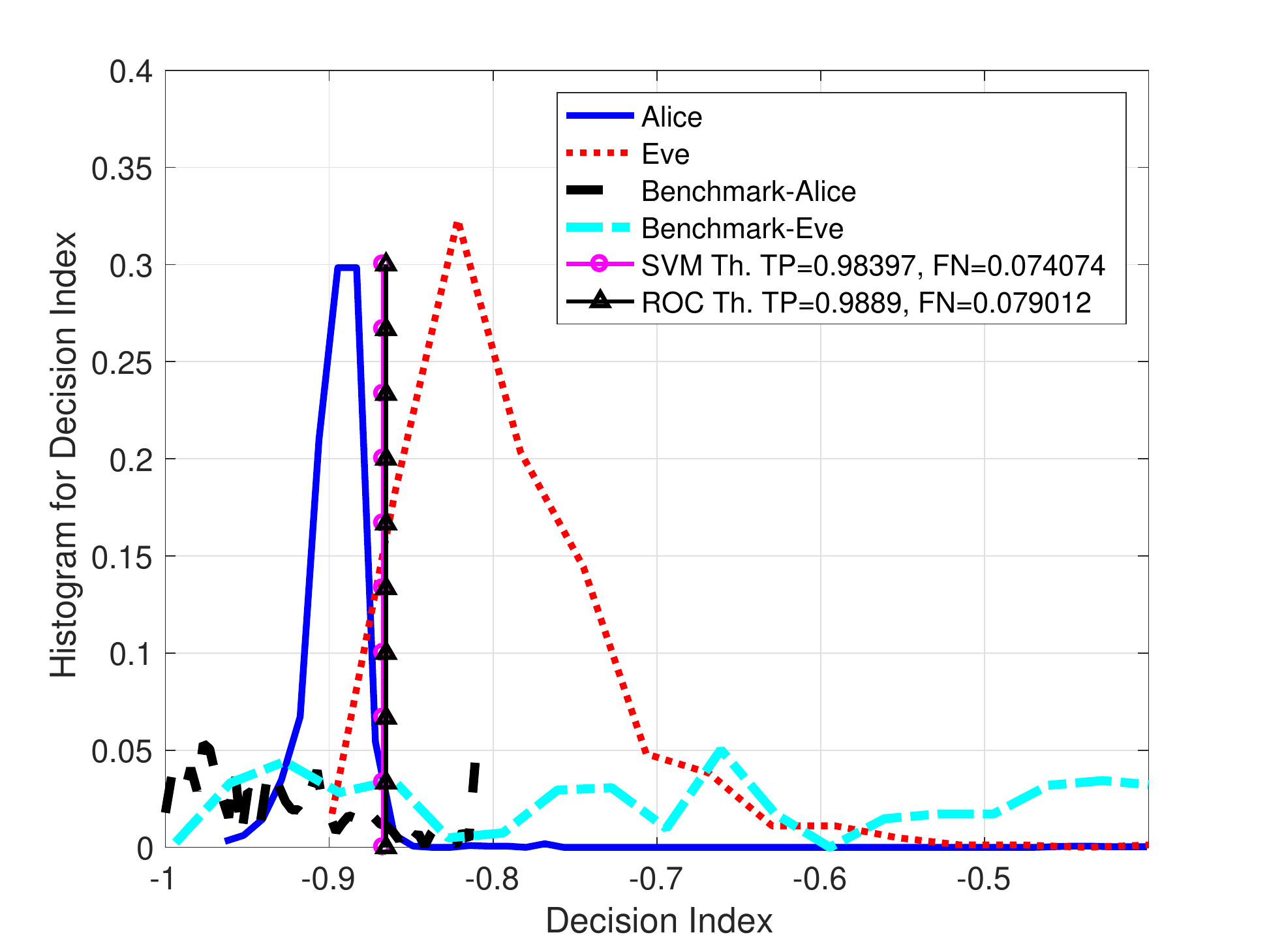}}%
\caption{Simulation results of decision index $\Psi_{\phi}$ from \eqref{e:threshold} for an advanced attacker.}
\label{f:hist}
\end{figure*}
\begin{figure*}[t]
       \centering
       \includegraphics[trim={0 40mm 0 50mm},clip,width=.85\columnwidth]{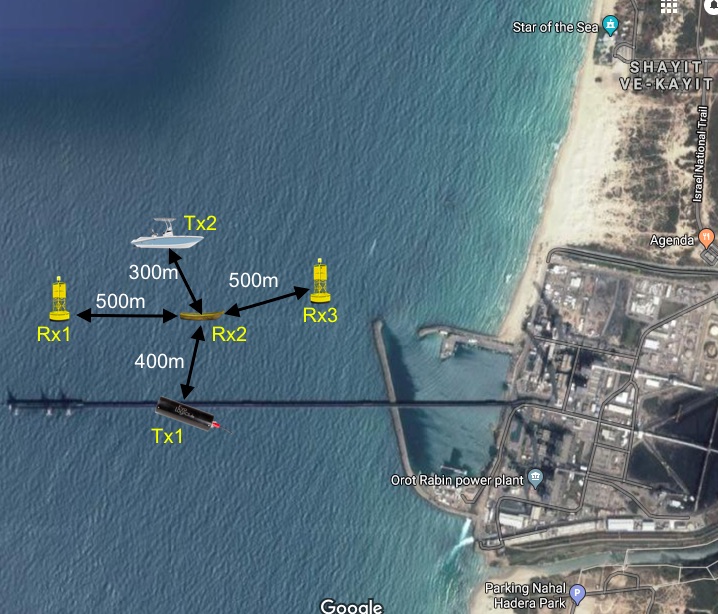}
       \hspace{5mm}
       \includegraphics[trim={0 0 0 9.5mm},clip,width=.85\columnwidth]{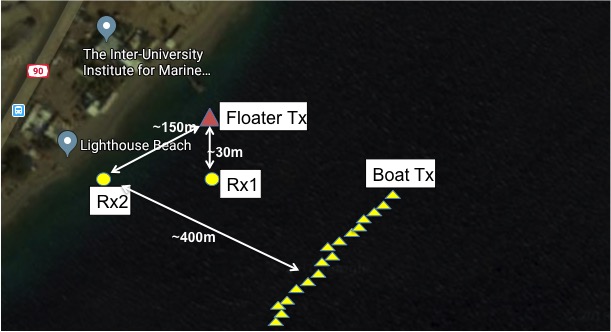}
       \caption{Setup of the Hadera (left) and Eilat (right) sea experiments.}
       \label{f:SeaTrialMap}
\end{figure*}

Next, we explore the performance of our authentication method against the more advanced TN-$x$ attacker models.
Recall that this requires the attacker to accurately estimate the bathymetry as well as the location of the legitimate node and the trusted node. We recall that we favor the attacker by assuming that it possesses perfect knowledge of the legitimate nodes and the trusted nodes' location, knowledge of the sound speed profile and of the sediments on the bottom (as is necessary to run the Bellhop propagation model), and slightly erroneous knowledge of the bathymetry.

\begin{figure*}[t]
\centering
    \subfloat[Relative RMS delay spread.]{\label{f:ExpDelaySpread}\includegraphics[width=0.85\columnwidth]{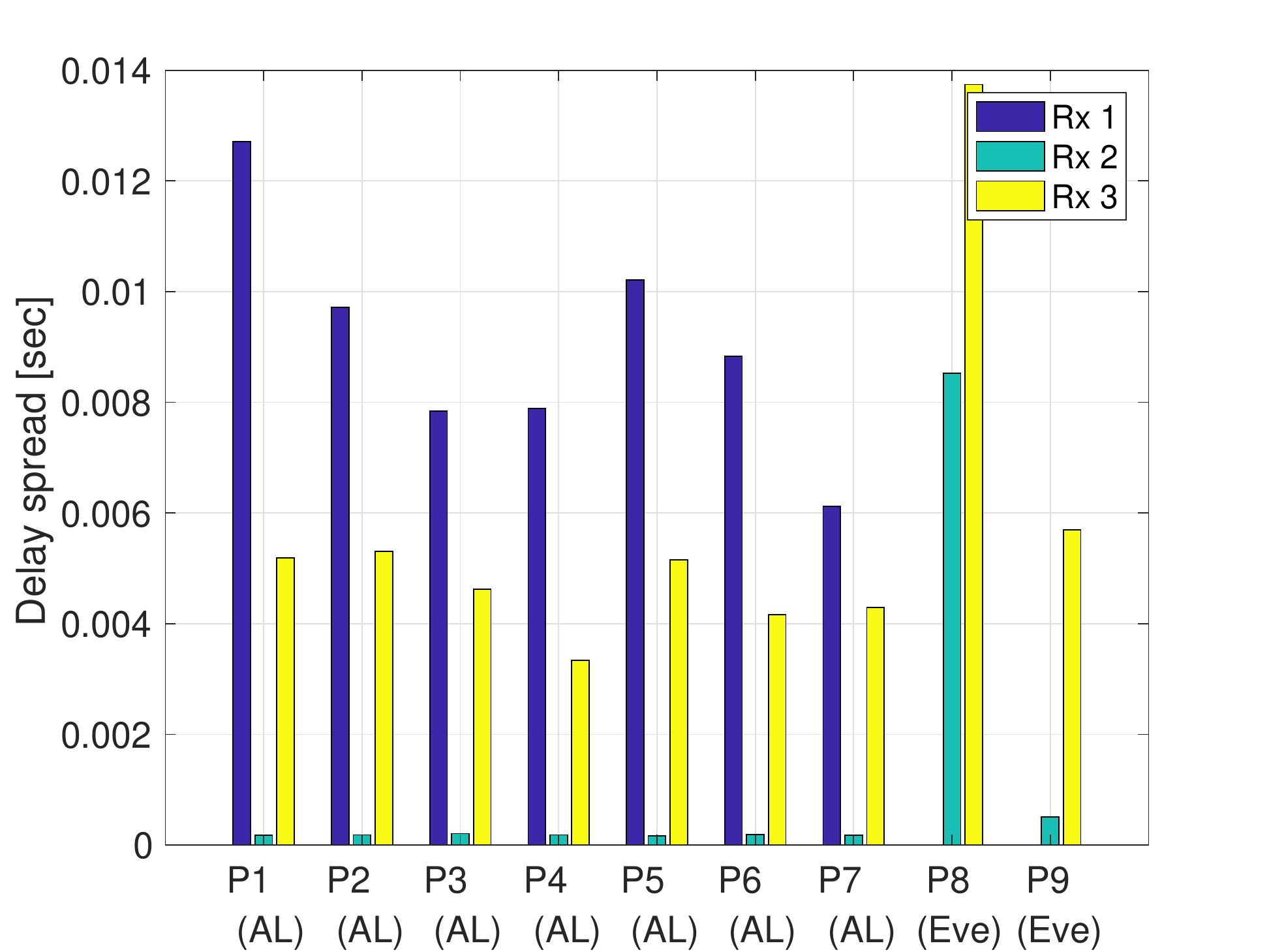}}\hspace{1em}%
    \subfloat[Number of taps.\label{f:ExpTapNum}]{\includegraphics[width=0.85\columnwidth]{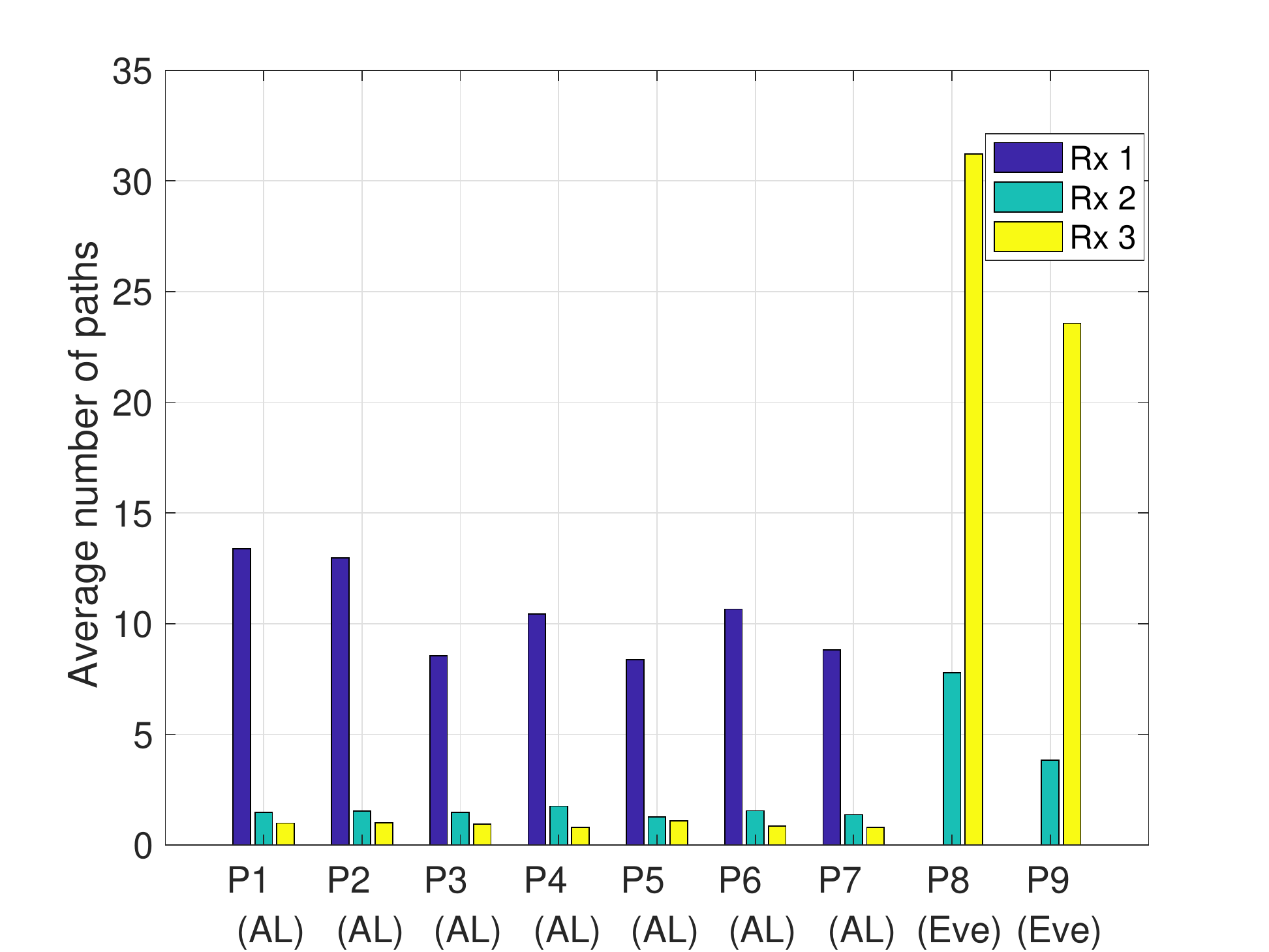}}\\[2.5mm]
    \subfloat[Average tap's power.\label{f:ExpTapPower}]{\includegraphics[width=0.85\columnwidth]{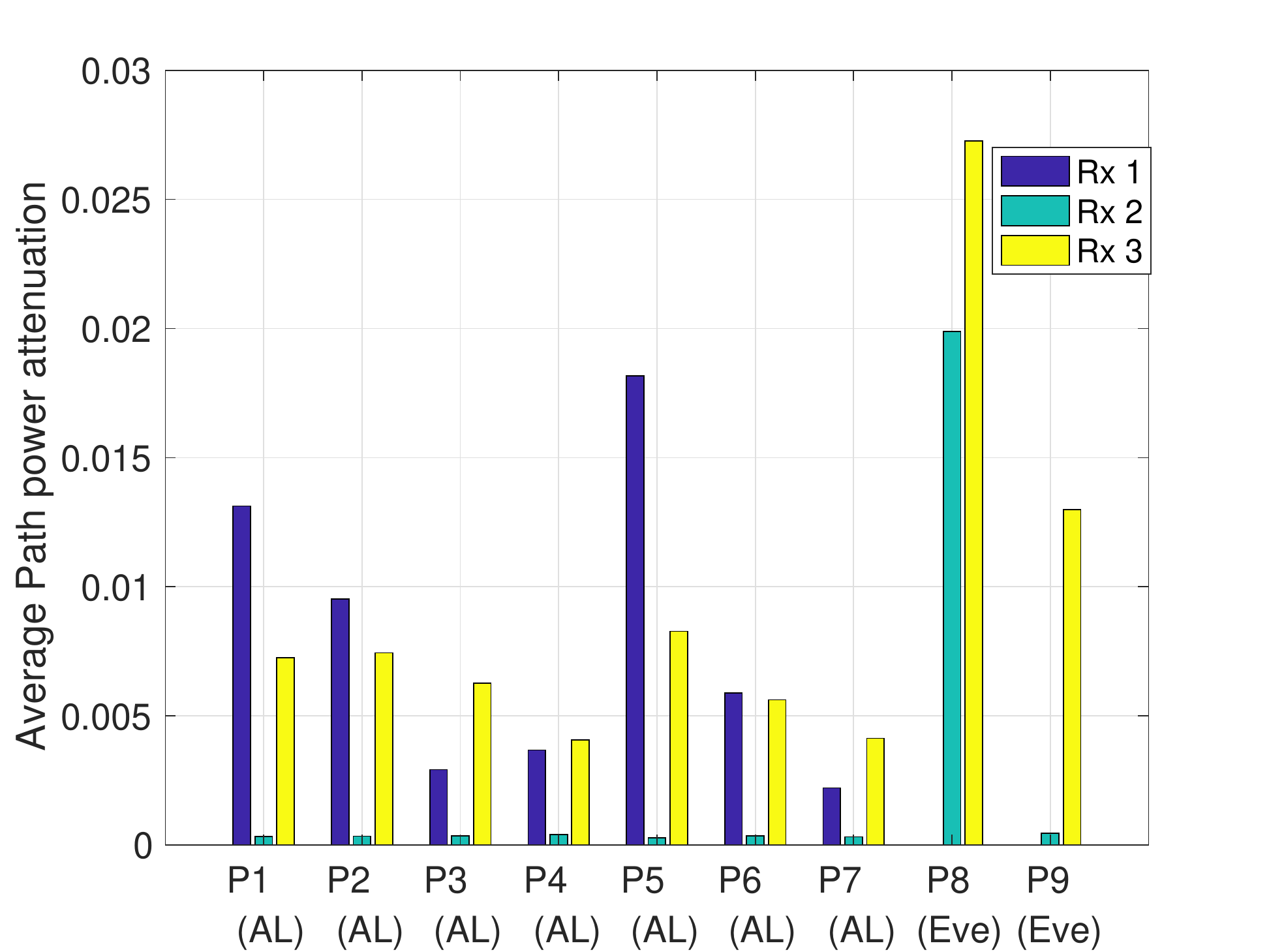}}\hspace{1em}%
    \subfloat[Smoothed received power level, $\alpha=0.5$.\label{subfig:smoothed}]{\label{f:ExpSmoothRL}\includegraphics[width=0.85\columnwidth]{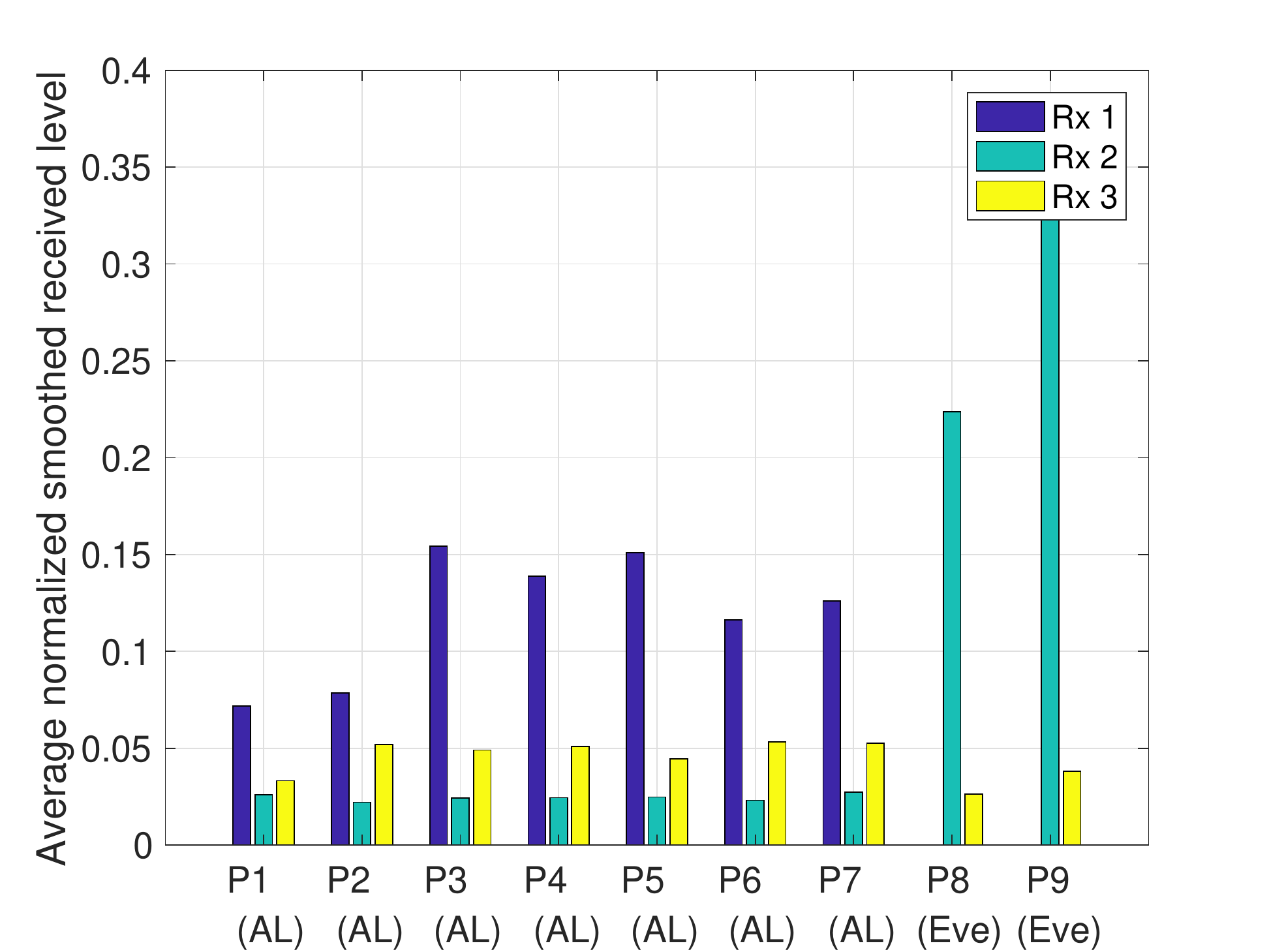}}%
    \caption{Average per-packet values of the authentication features from sea experiment Exp1 (Legitimate node: Tx 1).}
    \label{fig:simrexexp1}
\end{figure*}

We illustrate the authentication algorithm performance for the case where the attacker attempts to imitate the channel to a single trusted node (TN-1, Fig.~\ref{f:HistSimSmartEveNode1}); to two (TN-2, Fig.~\ref{f:HistSimSmartEveNode2}); to three (TN-3, Fig.~\ref{f:HistSimSmartEveNode3}); and to all four trusted nodes (TN-4, Fig.~\ref{f:HistSimSmartEveNode4}). Since the packets arrive almost simultaneously to all of the trusted nodes, the last three cases of TN-2, TN-3, and TN-4 are infeasible without an advanced and large transmission array. Still, we evaluate them to show that our approach remains robust even under largely unrealistic attacker models. For all cases, we also give the ROC performance as obtained by the statistical approach and by the SVM approach. In each caption, we provide the accuracy obtained through the SVM approach. By comparing the empirical distribution curves of the decision index for the two classes in Figs.~\ref{f:HistSimulationnaive} and~\ref{f:HistSimulationnaive}, we observe that the two classes become expectedly less distinct if the attacker imitates the channel to a higher number of trusted nodes. A similar, but less distinctive trend, is also observed for the benchmark scheme. For our approach , the results show that, even for the advanced TN-4 attacker, 92\% of the packets are still classified correctly. An even better authentication accuracy of 96\% is achieved for the more realistic TN-1 case. 

\section{Sea Experiment}\label{sec:exp}

In this section, we describe the setup and results obtained from two sea experiments aimed at demonstrating the effectiveness of our authentication method under real sea conditions. In particular, we show the validity of our assumptions regarding the distribution model, and measure the effectiveness to separate the legitimate and the attacker through the decision index~\eqref{e:threshold}. One experiment, \textit{Exp1}, was conducted in May 2017 near the Hadera coal pier in northern Israel, and the second, \textit{Exp2}, at the Inter-University Institute for marine sciences in Eilat, Israel. Exp1 enabled performance evaluation in a larger network, whereas Exp2 had a longer duration, and thus better reflects the variations of the UWAC channel over time.

\subsection{Setup} \label{sec:expsetup}
The setup of the sea experiments is illustrated in Fig.~\ref{f:SeaTrialMap}. Exp1 included two transmitters (station Tx1 on a pier that stretches 2~km into the sea, and station Tx2 on the larger boat), and three receivers (station Rx2 on the smaller boat, and stations Rx1 and Rx3 on floating buoys). The transmitters were placed roughly 1000~m apart, and the three receivers were deployed along a line with spacing of roughly 500~m. The distances between the receiver on the smaller boat and the boat and pier transmitters were roughly 300~m and 400~m, respectively. Due to the pier's thick stone columns, the transmissions of Tx1 were not received by Rx1. Exp2 included two transmitters, one on a buoy and one deployed from a boat, and two receivers, both deployed from anchored buoys. In both experiments, the water depth was about 25~m, and the modems and receivers were deployed at a depth of about 10~m. Since the area in Exp2 is a protected reef, all except the Boat Tx nodes were moored 
by scuba divers. During both experiments, the sea was rough, with a sea level of 3 and waves rising to over 1.5~m.

In both experiments, we used EvoLogics software-defined S2CR 7/17 modems to transmit from both transmitters packets of 100 (Exp1) and 60 (Exp2) chirp-modulated symbols of 10~ms in the frequency band between 7~kHz and 17~kHz at the source level of roughly 175~dB re 1$\mu$Pa@1m. The receivers employed a Cetacean CR1 hydrophone, and recorded continuously throughout the experiment. Analysis was performed off-line. In Exp1, we tested two configurations: one where node~1 served as the legitimate node and transmitted seven packets followed by the transmission of two packets from node~5, which served as the attacker, and vice versa in the second configuration. Exp2 was longer, as we let both the legitimate node (Floater Tx) and the attacker (Boat Tx) transmit sequences of 10 packets one after the other, for a total of 50 packets each. To allow accurate channel estimation, we used a guard interval of 0.1~s between each symbol, and set a gap that stretched from 10~s to 60~s between consecutive packets from the same transmitter. Exp1 lasted for 5~min for each configurations, while Exp2 lasted for 70~min. This long time span allowed capturing instantaneous channel variations, and testing our assumption in the realistic scenario of drifting nodes.    

As in the simulation discussed in  Section~\ref{sec:simres} where we did not consider the channels between nodes and sink, also in the experiments we 
collect data directly from the trusted nodes and process it off-line. The choice of channel features used for authentication is based on our preliminary analysis in Section~\ref{sec:param}, and includes the number of taps~\eqref{e:TapNum}, the relative RMS delay spread~\eqref{e:Delay}, and the smoothed received power level~\eqref{e:smooth}. Our results show that, to extract the largest degree of diversity from this scenario, including the average tap's power \eqref{e:TapPower} is beneficial due to the different link attenuations. Similar to our simulations, we estimated those channel features by thresholding the output of a normalized matched filter.

\subsection{Results} \label{sec:expres}

We start by showing the estimated values of the four authentication features from Exp1. In Fig.~\ref{f:ExpDelaySpread}, we show the average per-packet relative delay spread values from the three receivers. We observe that packets 8 and 9 from the attacker (Eve) are well distinguished from packets 1-7 from the legitimate node (AL). A similar effect is shown in Figs.~\ref{f:ExpTapNum} and \ref{f:ExpTapPower} for the average per-packet number of taps and the average tap power. Instead, Fig.~\ref{f:ExpSmoothRL} shows the per-packet smoothed received power level, and we observe that at Rx3 little difference is sensed between Alice's and Eve's. However,
at Rx2 the same metric results in 
very different values for Alice and Eve, 
hence the effect on performance is limited.  

\begin{figure*}[t]
    \centering
    \subfloat[Hadera experiment\label{f:sea_exp}]{\includegraphics[width=0.85\columnwidth]{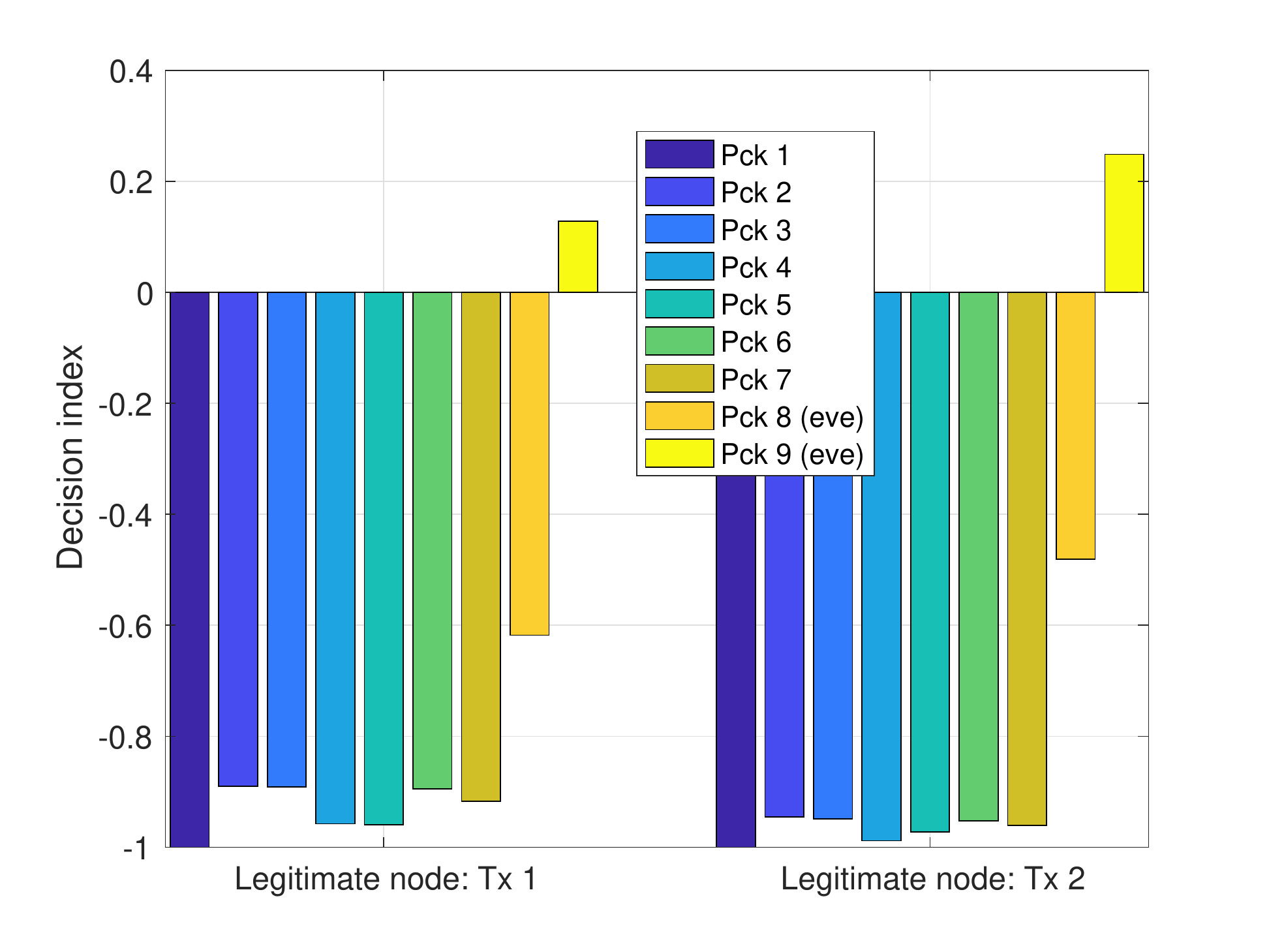}}
    \hspace{1em}
    \subfloat[Eilat experiment\label{f:sea_exp2}]{\includegraphics[width=0.85\columnwidth]{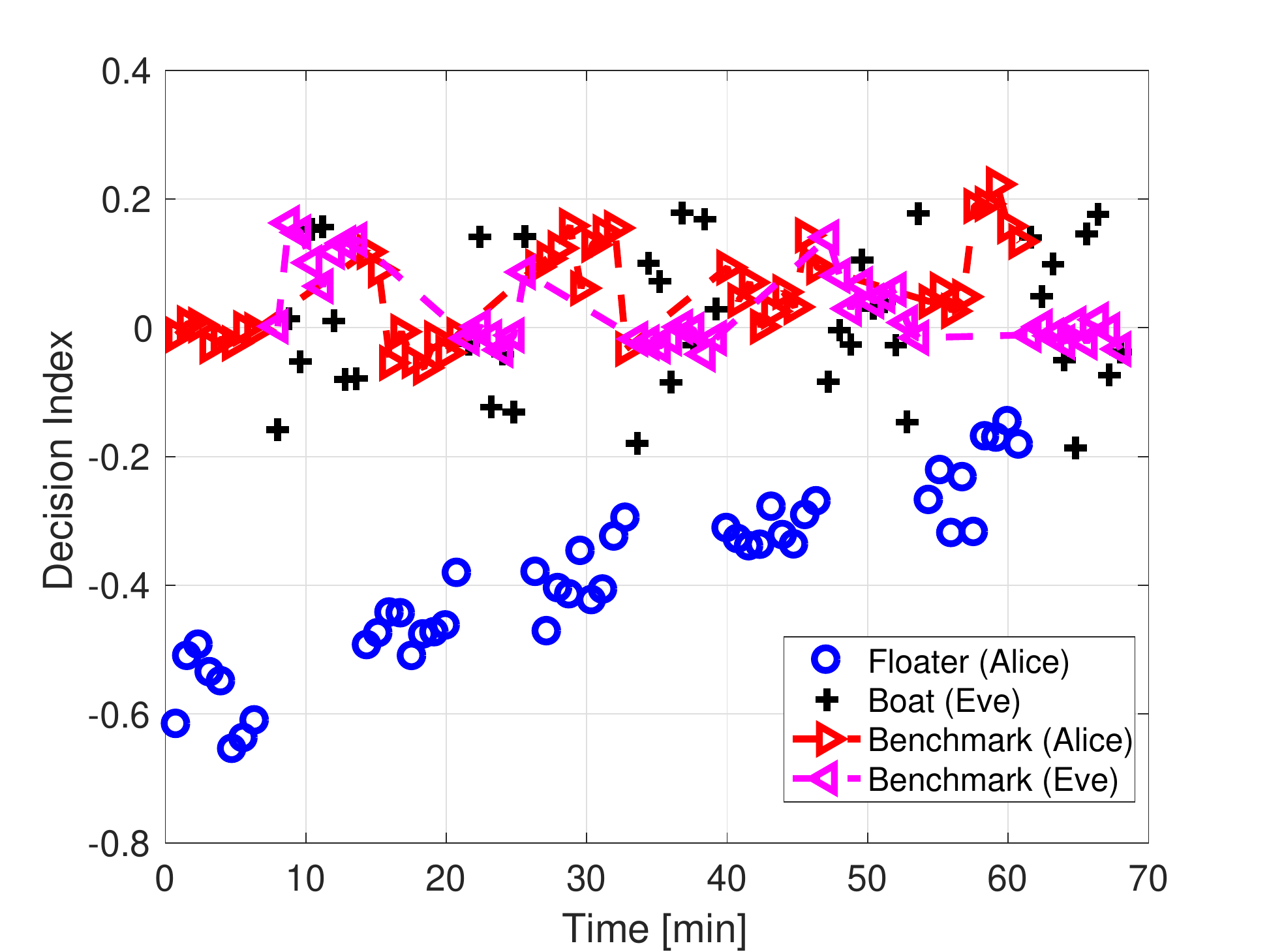}}%
    \caption{LLR $\Psi_{\phi}$ from the sea experiments, showing significant differences between Alice and Eve.}\vspace{2mm}
    \label{f:sea_exp_res}
\end{figure*}

Figs.~\ref{f:sea_exp} and~\ref{f:sea_exp2} show the LLR $\Psi_{\phi}$ obtained in Exp1 and Exp2 for all received packets. For Exp1, the LLRs from the legitimate transmitter are stable, whereas for Exp2 we observe a variation over time. This is because the longer duration of the experiment induces slow variations in the stability of the parameters $\hat{\omega}^m_{i,n}(\phi)$.
For both experiments $\Psi_{\phi}$ for the legitimate node's packets are well distinguished from the attacker's packets. In Fig.~\ref{f:sea_exp2} we also compare our results with those of the benchmark scheme. As expected, in realistic conditions the underwater channel changes over time, hence the decision index computed by the benchmark cannot distinguish between the legitimate node's and the attacker's packets. Still, as the distribution of the channel features remains relatively stable, our approach successfully discriminates authentic packets.

\begin{figure*}[t]
    \centering
    \subfloat[Hadera experiment.\label{fig:exp_param_imp_hadera}]{\includegraphics[width=0.85\columnwidth]{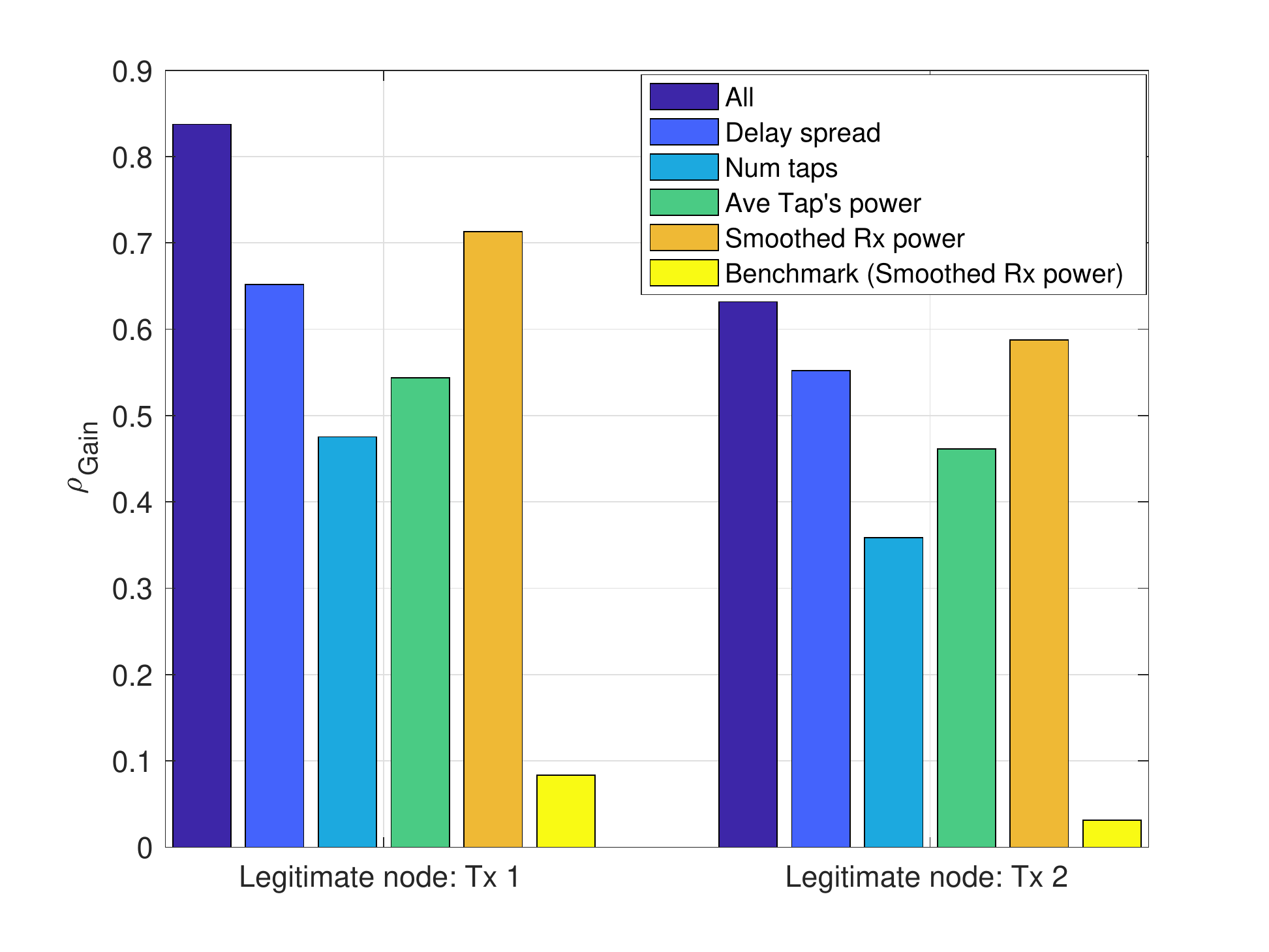}}
    \hspace{1em}
    \subfloat[Eilat experiment.\label{fig:exp_param_imp_eilat}]{\includegraphics[width=0.85\columnwidth]{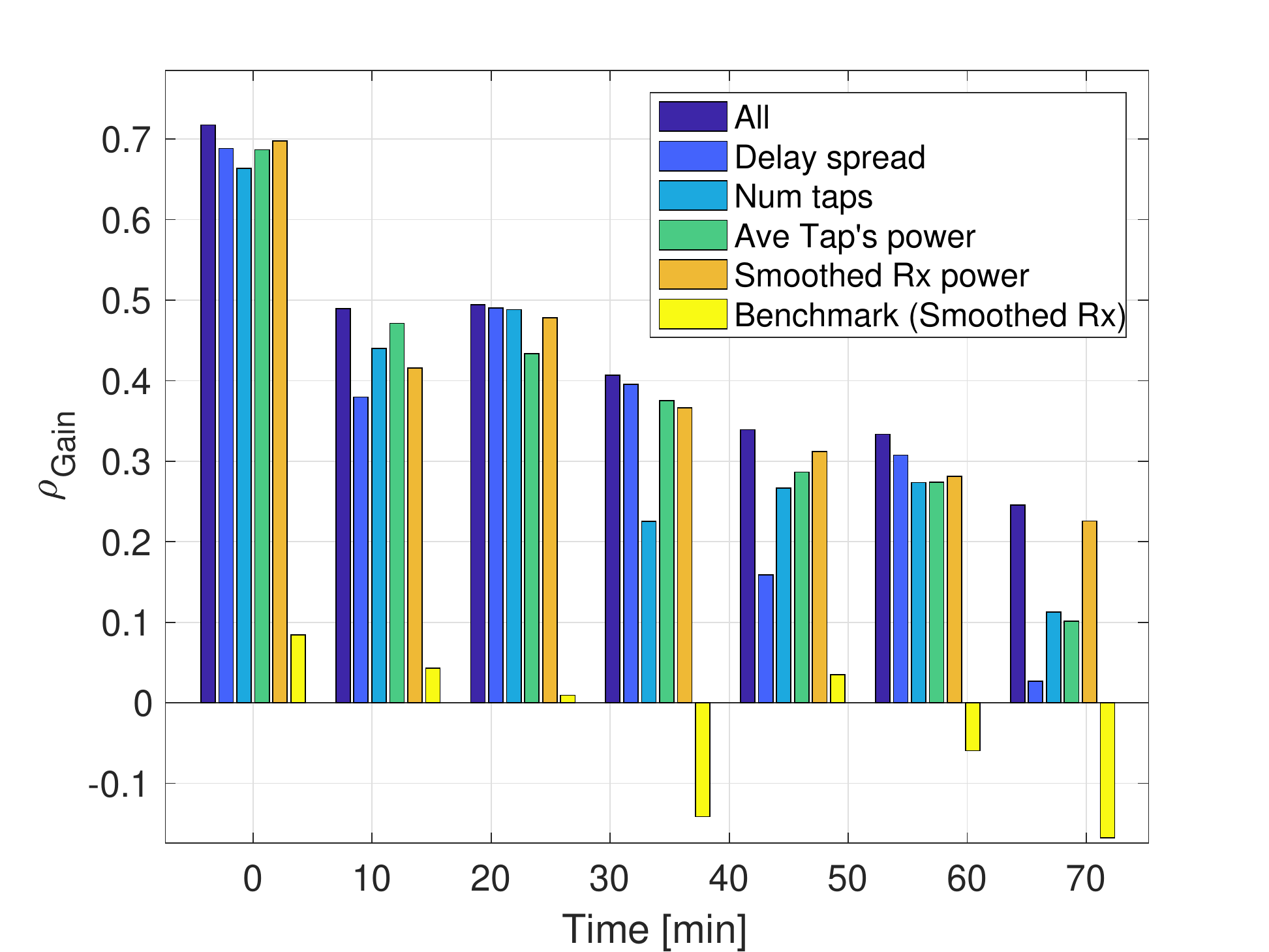}}
    \caption{Importance of each channel feature used for authentication in (a) the Hadera experiment and (b) the Eilat experiment.}\vspace{1.32mm}
    \label{fig:exp_param_imp}
\end{figure*}

In Fig.~\ref{fig:exp_param_imp}, we show the difference (or gain), $\rho_{\rm Gain}$, between the average $\Psi_{\phi}$ for the legitimate node's packets and the average $\Psi_{\phi}$ for the attacker's packets. For Exp1, the average was calculated for all packets, while for Exp2, due to the time variation observed in Fig.~\ref{f:sea_exp2}, we averaged groups of packets received over time windows of 10~minutes. To explore the contribution of each channel feature to the decision index, we show $\rho_{\rm Gain}$ when using each of the channel features alone and all the features together. For both experiments, the results of our method suggest that the conjunction of all features achieves higher gains and that, although the smoothed received power showed high gains, no single channel feature yields significantly higher benefits than the others. Fig.~\ref{fig:exp_param_imp} also shows $\rho_{\rm Gain}$ for the benchmark scheme applied to the smoothed received power. The small and sometimes negative gain reflects that, due to the channel's time variation, the benchmark is unable to leverage the diversity offered by the channel features and authentication fails. By way of contrast, the results of our method show a significant gain, and thus good authentication performance, even after 70~min of consecutive transmissions. Finally, we remark that, as shown in Fig.~\ref{f:sea_exp2}, $\rho_{\rm Gain}$ decreases over time, which implies on the need to re-estimate the distribution parameters of each features periodically, although this is only needed once every several tens of minutes, confirming the robustness of our scheme.

\section{Conclusions}\label{sec:conc}

In this paper, we presented a new cooperative message authentication method for underwater acoustic networks, where a set of trusted nodes helps a sink node determine whether a received message was sent by a legitimate transmitter or by an attacker. Our method leverages the strong spatial dependency and time-invariance of the distribution of the underwater acoustic channel features in order to perform authentication based on the computation of a decision index. This operation is performed by a set of trusted nodes acting in a distributed fashion, whereas a sink node fuses the beliefs of the trusted nodes and makes a final decision. No feedback to the trusted nodes is required for this.
Extensive model-based numerical simulations, as well as demonstrations in two sea experiments prove the effectiveness of our scheme. Further work will extend our approach to intruder detection.

\appendix

\label{parest}

In the following, we focus on a single channel feature $i$ and a single node $n$, and therefore drop the indices $i,n$ from all notation. To estimate $\theta(p)$, we use the expectation maximization (EM) algorithm~\cite[Ch.~17]{Mudd_EM} to update the estimate of $\theta$ iteratively. Denote $\tilde{\theta}(p+1)=\big(\tilde{\omega}^m(p+1), \tilde{k}^m(p+1)\big)$ as the estimate obtained at the end of iteration $p$.  

With the constraint that all estimates within one of the previously received packets $f=1,\ldots,\phi$ must be assigned to the same hypothesis ($m=0,1$), we define $y(f) \in \{0,1\}$ as the label assigned to packet $f$. Let $\mathcal Y(\phi) = \{y(t), t \in \mathcal{T}_f\}$ be the set of estimated labels for feature $i$ at node $n$ up to packet $\phi$, and $\bm{Y}(f) = \{y(t), t \in \mathcal{T}_f, f \leq \phi\}$. Let $\bar{{\cal X}}(f)$ be a subset of ${\cal X}(\phi)$ including only those measurements $x$ that are obtained from packet $f=\{0, \ldots,\phi\}$. The expectation of the log-likelihood function at iteration $p$, is
\begin{equation}
\begin{split}
L\big(\theta&|\tilde{\theta}(p)\big) = {\mathbb E}\left[\log\big({\mathbb P}({\mathcal X}(\phi),{\mathcal Y}(\phi)|\theta)\big)|\bar{{\cal X}}(\phi),\tilde{\theta}(p)\right]=\\
&=\sum_{m=0}^{1}\sum_{f=1}^{F}{\mathbb P}\big(y(f)=m|\bm{X}(f),\tilde{\theta}(p)\big)\cdot\\
&\bigg[\log \tilde{k}^m(p)+\sum_{x\in\bar{{\cal X}}(f)}\log p_{x|\omega^m}\big(x|\tilde{\omega}^m(p)\big)\bigg]\;,
\label{e:log}
\end{split}
\end{equation}
where $\log x$ is the natural logarithm of $x$ and the posterior is
\begin{equation}
\begin{split}
{\mathbb P}\big(y(f)&=m|\bar{{\cal X}}(f),\tilde{\theta}(p)\big)\!=\!
\frac{\tilde{k}^m(p){\mathbb P}\big({\cal X}(f)|\tilde{\omega}^m(p)\big)}{{\mathbb P}\big(\bar{{\cal X}}(f)|\tilde{\theta}(p)\big)}\\
&=\frac{\tilde{k}^m(p)\prod\limits_{x\in\bar{{\cal X}}(f)}p_{x|\omega^m}\big(x|\tilde{\omega}^m(p)\big)}
{\sum_{j=1}^{2}\tilde{k}^j(p)\prod\limits_{x \in\bar{{\cal X}}(f)}
p_{x|\omega^m}\big(x|\tilde{\omega}^j(p)\big)}\;.
\end{split}
\label{e:p_m}
\end{equation}
We maximize the second term of \eqref{e:log} to yield \cite{Diamant:LOS},
\begin{equation}
\tilde{k}^m(p+1)=\frac{1}{\phi}\sum_{f=1}^{\phi}{\mathbb P}(y(f)=m|\bar{{\cal X}}(f),\tilde{\theta}(p))\;.
\label{e:Est_k2}
\end{equation}

To obtain $\tilde{\omega}^m(p+1)$, we maximize the first term of \eqref{e:log}
\begin{equation}
\begin{split}
&\sum_{f=1}^{\phi}{\mathbb P}\big(y(f)=m|\bar{{\cal X}}(f),\tilde{\theta}(p)\big)\cdot\log {\mathbb P}\big(\bar{{\cal X}}(f)|\tilde{\omega}^m(p+1)\big)= \\
&= \sum_{f=1}^{\phi}{\mathbb P}\big(y(f)=m|\bar{{\cal X}}(f),\tilde{\theta}(p)\big)\cdot \\
&\cdot \!\!\!\sum_{x\in\bar{{\cal X}}(f)}\!\!\bigg[\log\tilde{\beta}^m(p+1)\!-\!\log\big(2\tilde{\sigma}^m(p+1)\big)- \\
&  - \log\Gamma\left(\frac{1}{\tilde{\beta}^m(p+1)}\right)\!-\!\left(\frac{|x-\tilde{\mu}^m(p+1)|}{\tilde{\sigma}^m(p+1)}\right)^{\tilde{\beta}^m(p+1)}\bigg].
\end{split}
\label{e:log_1}
\end{equation}
The result is obtained numerically by solving the following equation for $\tilde{\mu}^m(p+1)$:
\begin{equation}
\label{e:sol_a}
\begin{split}
    0= & \sum_{f=1}^{\phi}{\mathbb P}\big(y(f) = m|\bar{{\cal X}}(f),\tilde{\theta}(p)\big) \times \\
     & \times  \sum_{x\in\bar{{\cal X}}(f)}\big(x-\tilde{\mu}^m(p+1)\big)\big(|x-\tilde{\mu}^m(p+1)|\big)^{\tilde{\beta}^m(p)-2}\;,
\end{split}
\end{equation}
and then by solving, for $\tilde{\sigma}^m(p+1)$,
\begin{equation}
\label{e:sol_b}
\begin{split}
    &0\!= \sum_{f=1}^{\phi}{\mathbb P}\big(y(f) \!=\! m|\bar{{\cal X}}(f),\tilde{\theta}(p)\big)\\
    &\sum_{x\in\bar{{\cal X}}(f)}\!\!\!\left(-\frac{1}{\tilde{\sigma}^m(p+1)}+
   \frac{\tilde{\beta}^m(p)\big(|x-\tilde{\mu}^m(p+1)|\big)^{\tilde{\beta}^m(p)}}
  {\left(\tilde{\sigma}^m(p+1)\right)^{\tilde{\beta}^m(p)-1}} 
 \right).
 \end{split}
\end{equation}

\noindent Finally, for $\tilde{\beta}^m(p+1)$, we solve ($d(\cdot)$ is the digamma function)
\begin{equation}
\label{e:sol_c}
\begin{split}
&0=\sum_{f=1}^{\phi}{\mathbb P}\big(y(f) = m|\bar{{\cal X}}(f),\theta(p)\big)\left[\frac{1}{\tilde{\beta}^m(p+1)}+\frac{d\left(\frac{1}{\tilde{\beta}^m(p+1)}\right)}{(\tilde{\beta}^m(p+1))^2}\right.\\
&+\sum_{x\in\bar{{\cal X}}(f)}\left.\log\left(\frac{|x-\tilde{\mu}^m(p+1)|}{\tilde{\sigma}^m(p+1)}\right)\cdot
\left(\frac{|x-\tilde{\mu}^m(p+1)|}{\tilde{\sigma}^m(p+1)}\right)^{\tilde{\beta}^m(p+1)}\right]\;.
\end{split}
\end{equation}

After the convergence of the EM procedure, we calculate the posterior ${\mathbb P}\big(y(f) = m|\bar{{\cal X}}(f),\tilde{\theta}(p+1)\big)$, and use it as a soft decision parameter to assign packet $f$ with set $m$. To consistently identify sets $m=\{0,1\}$ between the different trusted nodes, for the first received packet, we determine $m=1$ as the set $m$ that yields the highest posterior.

To initialize the EM, we use the k-means algorithm~\cite[Section~33.7]{DJCMacKay2005} to cluster ${\mathcal X}(\phi)$ into two groups. For each group $m$, we evaluate $\tilde{\omega}^m(0)$ using the following statistics for \ac{PDF} (\ref{e:distribution})
\begin{subequations}
\label{e:Relation}
    \begin{align}
     &{\mathbb E}\left[{\mathcal X}(\phi)\right]=\tilde{\mu}^m(0)\\
     &{\mathbb E}\left[|{\mathcal X}(\phi) - \tilde{\mu}^m(0)|^2\right] =\frac{\tilde{\sigma}^{m}(0)^2 \, \Gamma\left(\frac{3}{\tilde{\beta}^m(0)}\right)}{\Gamma\left(\frac{1}{\tilde{\beta}^m(0)}\right)}\\ 
  &\mathrm{Kurtosis} = \frac{{\mathbb E}[({\mathcal X}(\phi) - \tilde{\mu}^m(0))^2]}{({\mathbb E}[({\mathcal X}(\phi) - \tilde{\mu}^m(0))^2])^2} =\nonumber\\
  &=\frac{\Gamma\left(\frac{5}{\tilde{\beta}^m(0)}\right)\Gamma\left(\frac{1}{\tilde{\beta}^m(0)}\right)}{\Gamma\left(\frac{3}{\tilde{\beta}^m(0)}\right)^2}-3\;.
    \end{align}
\end{subequations}
where the above mean, variance, and Kurtosis are calculated using standard methods. Similarly, the prior $\tilde{k}^m(0)$ is estimated as the fraction of the elements of ${\cal X}(\phi)$, which is part of the $m$th cluster after the k-means algorithm.

\bibliographystyle{IEEEtran}
\bibliography{IEEEabrv,SecurityBib}

\vfill

\begin{IEEEbiography}[{\includegraphics[width=1in,height=1.25in,clip,keepaspectratio]%
{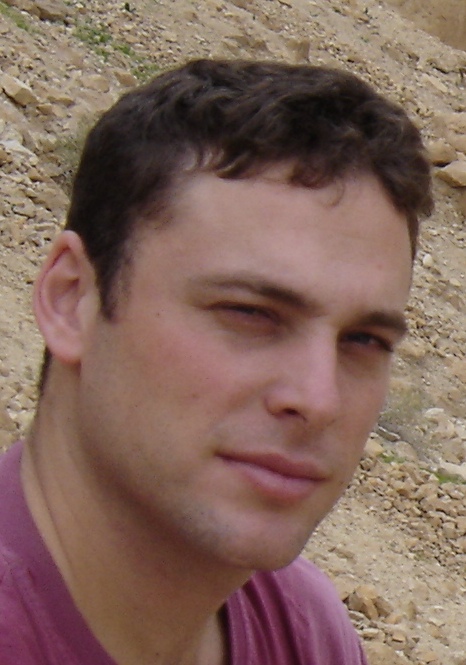}}]{\bf Roee Diamant} (S'09--M'15--SM'17) received the B.Sc.\ and M.Sc.\ degrees from Technion--Israel Institute of Technology, in 2002 and 2007, respectively,
and the Ph.D.\ degree from the Department of Electrical and Computer Engineering, The University of
British Columbia, in 2013. From 2001 to 2009, he was with Rafael Advanced Defense Systems, Israel,
as the Project Manager and Systems Engineer, where he developed a commercial underwater modem with
network capabilities. In 2015 and 2016, he was a Visiting Professor with the University of Padova,
Italy. He leads the underwater Acoustic and Navigation Laboratory as an Assistant Professor at the Department of Marine Technologies, University of Haifa. His research interests include underwater acoustic communication, underwater navigation, object detection, and classification. He is the Coordinator of the EU H2020 Project SYMBIOSIS (BG-14 track). In 2009, he received the Israel Excellent Worker First Place Award from the Israeli Presidential Institute. In 2010, he received the NSERC Vanier Canada Graduate Scholarship. He has received three best paper awards, and serves as an Associate Editor for the \textsc{IEEE Journal of Oceanic Engineering}.
\end{IEEEbiography}

\begin{IEEEbiography}[{\includegraphics[width=1in,height=1.25in,clip,keepaspectratio]%
{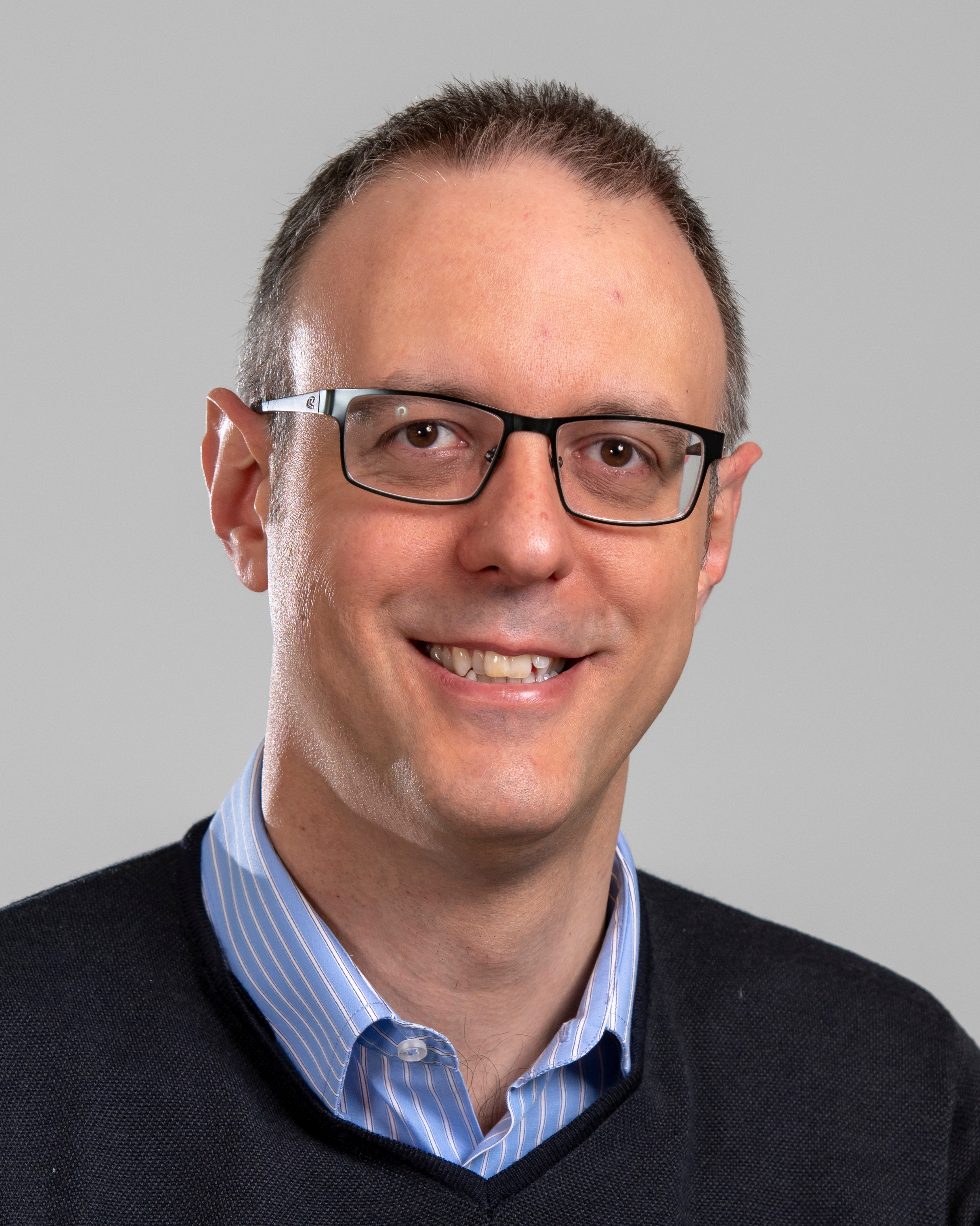}}]{\bf Paolo Casari} (S'05--M'08--SM'13) received the PhD in Information Engineering in 2008 from the University of Padova, Italy. After being on leave at the Massachusetts Institute of Technology in 2007, his research progressively focused on underwater communications and networks. He collaborated to several funded projects including CLAM (EU-FP7), RACUN (EDA), and several US ARO, ONR and NSF initiatives. He is the PI of the NATO SPS project ThreatDetect, and the scientific coordinator of the EU H2020 RECAP and SYMBIOSIS projects. Previously, he was appointed Technical Manager of the NAUTILUS and WISEWAI projects. In 2015, he joined the IMDEA Networks Institute, Madrid, Spain, where he leads the Ubiquitous Wireless Networks group. His research interests include many aspects of underwater communications, such as channel modeling and simulation, network protocol design and evaluation, localization, and field experiments. He was a co-recipient of the IEEE WPNC 2017 best paper award. 
He serves on the editorial boards of the \textsc{IEEE Transactions on Mobile Computing} and of the \textsc{IEEE Transactions on Wireless Communications}, and regularly collaborates with the organizing committee of several conferences. He has been guest editor of a special section of \textsc{IEEE Access} on Underwater Wireless Communications and Networking, and of a special issue of the \emph{Hindawi Journal of Electrical and Computer Engineering} on Underwater Communications and Networking. 
\end{IEEEbiography}

\enlargethispage{-8cm}
\begin{IEEEbiography}[{\includegraphics[width=1in,height=1.25in,clip,keepaspectratio]%
{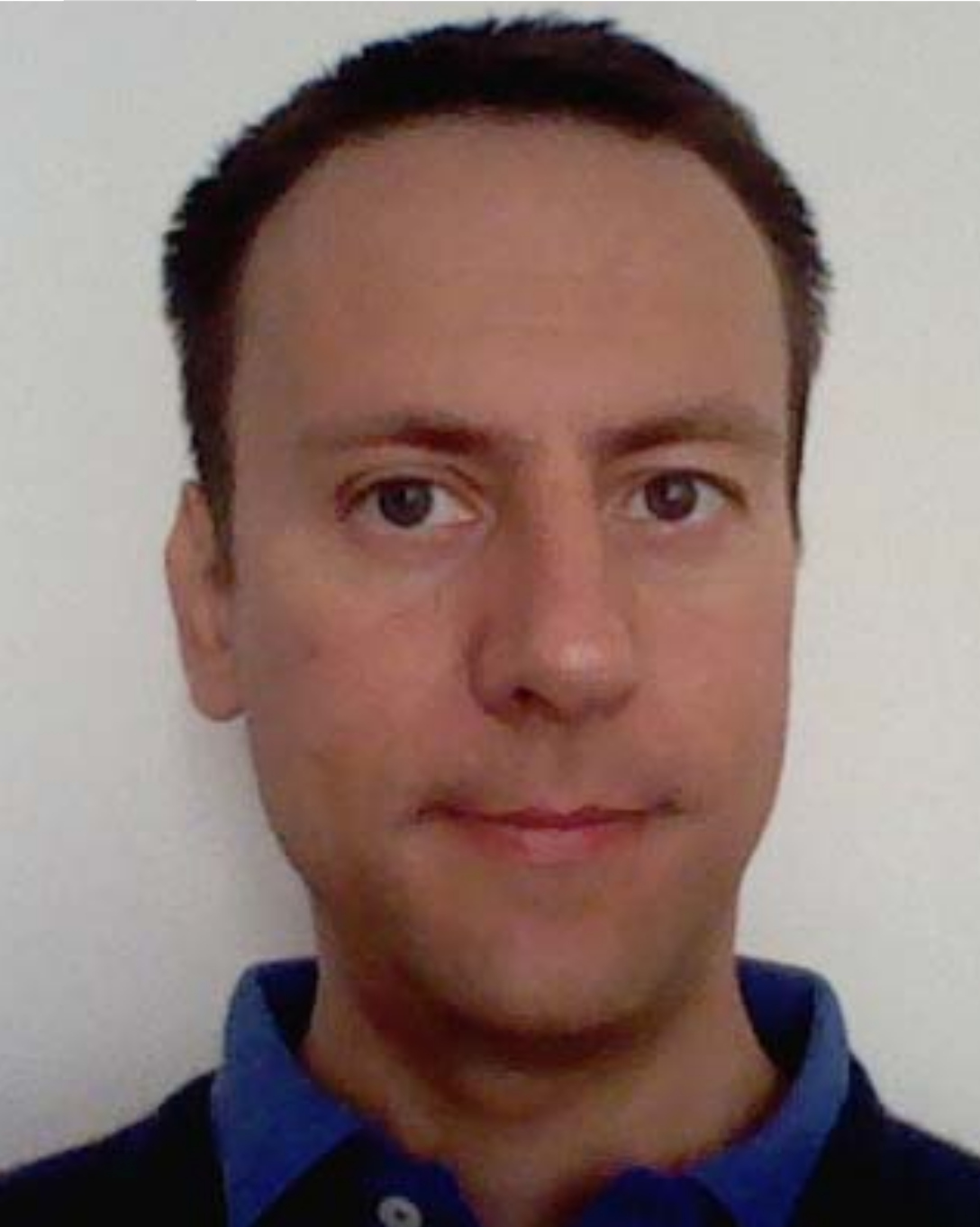}}]{\bf Stefano Tomasin} (S'99--M'03--SM'11) was born in Venice, Italy.
He received the Ph.D.\ degree in telecommunications engineering from the University of Padova, Italy, in 2003. In 2002, he joined the University of Padova, where he is currently an Associate Professor.
He was on leave with Philips Research, Eindhoven, Netherlands, in 2002, Qualcomm Research Laboratories, San Diego, CA, USA, in 2004, Polytechnic University, Brooklyn, NY, USA, in 2007, and the Huawei Mathematical and Algorithmic Sciences Laboratory, Boulogne-Billancourt, France, in 2014. His current research interests include physical layer security and signal processing for wireless communications, with application to 5th generation cellular systems. Since 2011, he has been an Editor of the \emph{EURASIP Journal of Wireless Communications and Networking}. From 2011 to 2017, he was an Editor of the \textsc{IEEE Transactions on Vehicular Technology}, and has been the Editor of the \textsc{IEEE Transactions on Signal Processing} since 2016.
\end{IEEEbiography}

\end{document}